\documentclass{aa}

\usepackage{enumerate}

\makeatletter
\g@addto@macro{\UrlBreaks}{\UrlOrds}
\makeatother
\usepackage{CJKutf8}

\newcommand{\hg}{H$\gamma$}
\newcommand{\hd}{H$\delta$}
\newcommand{\dfn}{D$_\mathrm{N}4000$}
\newcommand{\hda}{\hd$_\mathrm{A}$}

\newcommand{\sfr}{SFR$_\mathrm{100Myr}$}

\usepackage{import}
\usepackage{amsmath}
\usepackage{amssymb}
\usepackage{natbib}
\bibpunct{(}{)}{;}{a}{}{,}
\usepackage{lscape}
\usepackage{xcolor}
\usepackage[normalem]{ulem}
\usepackage{blindtext}
\usepackage{graphicx}
\usepackage[para]{threeparttable}
\usepackage{subfigure}  
\usepackage{txfonts}
\usepackage[colorlinks=true,linkcolor=blue,citecolor=blue,urlcolor=blue]{hyperref}
\begin{document} 

   \title{The LEGA-C galaxy survey: multiple quenching channels for quiescent galaxies at $z\sim1$}
   \titlerunning{Multiple quenching pathways for quiescent galaxies at $z\sim1$}
   
   \author{Angelos Nersesian\inst{1, 2}
          \and
          Yasha Kaushal\inst{3}
          \and
          Marco Martorano\inst{2}
          \and
          Arjen van der Wel\inst{2,4}
          \and 
          Po-Feng Wu\inst{5,6,7}
          \and 
          Rachel Bezanson\inst{3}
          \and
          Eric F. Bell\inst{8}
          \and
          Francesco D'Eugenio\inst{9,10}
          \and
          Anna R. Gallazzi\inst{11}
          \and
          Joel Leja\inst{12,13,14} 
          \and
          Stefano Zibetti\inst{11}
          \and 
          Sandro Tacchella\inst{9,10}
          }
   \institute{STAR Institute, Universit\'e de Li{\`e}ge, Quartier Agora, All\'ee du six Aout 19c, B-4000 Liege, Belgium\\
        \email{\textcolor{blue}{angelos.nersesian@uliege.be}}
    \and
        Sterrenkundig Observatorium Universiteit Gent, Krijgslaan 281 S9, B-9000 Gent, Belgium
   \and
        Department of Physics and Astronomy and PITT PACC, University of Pittsburgh, Pittsburgh, PA 15260, USA
    \and
        Max-Planck Institut f\"{u}r Astronomie K\"{o}nigstuhl, D-69117, Heidelberg, Germany
    \and
        Institute of Astrophysics, National Taiwan University, Taipei 10617, Taiwan
    \and
        Department of Physics and Center for Theoretical Physics, National Taiwan University, Taipei 10617, Taiwan
    \and
        Physics Division, National Center for Theoretical Sciences, Taipei 10617, Taiwan
    \and 
        Department of Astronomy, University of Michigan, 1085 South University Avenue, Ann Arbor, MI 48109, USA
    \and
        Kavli Institute for Cosmology, University of Cambridge, Madingley Road, Cambridge, CB3 0HA, UK
    \and 
        Cavendish Laboratory - Astrophysics Group, University of Cambridge, 19 JJ Thomson Avenue, Cambridge, CB3 0HE, UK 
    \and 
        Osservatorio Astrofisico di Arcetri, Largo Enrico Fermi 5, I-50125 Firenze, Italy
    \and 
        Department of Astronomy and Astrophysics, 525 Davey Lab, The Pennsylvania State University, University Park, PA 16802, USA
    \and
        Institute for Gravitation and the Cosmos, The Pennsylvania State University, University Park, PA 16802, USA
    \and
        Institute for Computational and Data Sciences, The Pennsylvania State University, University Park, PA 16802, USA   
    }

   \date{Received 17 September 2025 / Accepted 9 December 2025}

 
  \abstract
   {}
   {We analyzed the sizes and star-formation histories (SFHs) of 2908 galaxies with $M_\star \geq 10^9$~M$_\odot$ at $0.6 < z < 1.0$, drawn from the Large Early Galaxy Astrophysics Census (LEGA-C) survey. The goal is to investigate the connection between galaxy sizes with SFH, stellar age, and metallicity.}
   {SFHs were derived with {\tt Prospector} by fitting the high signal-to-noise, high spectral resolution spectroscopy drawn from the LEGA-C DR3 together with the broadband photometry from the UltraVISTA catalog. Galaxy sizes were measured by fitting a 2D S{\'e}rsic profile to the \textit{HST} ACS~F814W images.}
   {We find diverse SFHs and quenching timescales ($\tau_\mathrm{q}$). The main quiescent population quenched over $\tau_\mathrm{q}=1.23\pm0.04$~Gyr, whereas compact post-starburst galaxies (PSBs) quenched much faster, $\tau_\mathrm{q}=0.13\pm0.03$~Gyr. At fixed stellar mass, smaller quiescent galaxies quenched more rapidly than larger ones; at fixed size, the dependence on stellar mass is weak. Larger quiescent galaxies are marginally younger, quenched more slowly, and have near-solar metallicities, while compact quiescent galaxies are older, metal-rich, and quenched faster. PSBs formed half their mass later ($z_\mathrm{form}\sim1.9$) and quenched on the shortest timescales. The general trends with galaxy size, $Z_\star$, and $z_\mathrm{form}$ for the quiescent populations remain consistent regardless of the method used to derive the stellar properties.}
   {We conclude that compact quiescent galaxies are consistent with both early, moderately fast quenching and with more rapid, late quenching. While this may suggest the existence of multiple quenching channels, our data are also compatible with a continuous distribution of quenching timescales. These findings suggest that different physical mechanisms may drive quenching across galaxy populations, potentially leading to similar morphological outcomes despite differing evolutionary histories.}


   \maketitle
%

\section{Introduction}

Galaxies exhibit a large diversity in morphology, chemical composition, and star-formation histories (SFHs), reflecting the multitude of evolutionary pathways they can follow. A key question in galaxy evolution is understanding the processes that drive the cessation of star formation, commonly referred to as quenching. Quenching in massive galaxies is thought to result from a combination of mechanisms, including the depletion of cold gas reservoirs \citep{Springel_2005ApJ...620L..79S, Zhang_2019ApJ...884L..52Z}, virial shock heating in massive halos \citep{Birnboim_2003MNRAS.345..349B, Birnboim_2007MNRAS.380..339B, Keres_2005MNRAS.363....2K, Dekel_2006MNRAS.368....2D, Mitra_2015MNRAS.452.1184M}, and feedback processes such as supernova explosions or active galactic nuclei (AGN) activity that inhibit gas cooling \citep[e.g.,][]{Crenshaw_2003ARA&A..41..117C, Best_2005MNRAS.362...25B, Croton_2006MNRAS.365...11C, Fabian_2012ARA&A..50..455F, Cheung_2016Natur.533..504C, Harrison_2017NatAs...1E.165H, Barisic_2017ApJ...847...72B, Terrazas_2017ApJ...844..170T, Henden_2018MNRAS.479.5385H, Semenov_2021ApJ...918...13S, Dome_2024MNRAS.527.2139D}. The efficiency of these mechanisms varies with galaxy mass, morphology, and environment, creating a complex interplay that shapes galaxy properties over time.

One aspect of this complexity is the relationship between galaxy sizes, SFHs, and stellar population properties. Observations show that quiescent galaxies in the local Universe are typically smaller than star-forming galaxies of the same stellar mass, following steeper size--mass relations \citep{Shen_2003MNRAS.343..978S}. This correlation is further complicated by evidence that the quenching process itself can modify galaxy structure \citep{McDermid_2015MNRAS.448.3484M, Wu_2018ApJ...868...37W, Hamadouche_2022MNRAS.512.1262H}. For example, energy injection from feedback mechanisms can suppress star formation without drastically altering stellar distributions \citep{Dekel_2006MNRAS.368....2D, Fabian_2012ARA&A..50..455F, Terrazas_2017ApJ...844..170T}, whereas rapid quenching due to starbursts, gas-rich galaxy mergers, or disk instabilities can lead to galaxies with dense stellar cores \citep{Springel_2005ApJ...620L..79S, Snyder_2011ApJ...741...77S, Zolotov_2015MNRAS.450.2327Z, Tacchella_2016MNRAS.458..242T, van_der_Wel_2025arXiv250902847V}. These processes imprint signatures of quenching events onto galaxy morphologies, making structural properties a valuable tracer of galaxy evolution.

Evidence from large extragalactic surveys as well as hydrodynamical simulations, further support differences in galaxy structure as a function of star formation. Studies in the local Universe show evidence of compact galaxies quenching faster than galaxies with an extended stellar disk \citep[e.g.,][]{McDermid_2015MNRAS.448.3484M, Barone_2020ApJ...898...62B}. Conversely, massive galaxies with an extended disk build up their stellar content slowly over a longer timescale \citep{Boissier_2016A&A...593A.126B, Zhu_2018MNRAS.480L..18Z}. Cosmological simulations also indicate that galaxies with extended stellar disks tend to quench more slowly compared to their compact counterparts \citep{Pillepich_2018MNRAS.473.4077P, Tacchella_2019MNRAS.487.5416T, Gupta_2021ApJ...907...95G, Park_2022MNRAS.515..213P}. 

Consequently, the connection between galaxy sizes and stellar population properties offers a unique insight into quenching pathways. At fixed stellar mass, larger quiescent galaxies tend to have younger stellar populations in their outskirts, consistent with size growth through processes like minor mergers \citep{Shankar_2010MNRAS.403..117S, Carollo_2013ApJ...773..112C}. Conversely, compact quiescent galaxies appear to quench earlier, resulting in older stellar populations \citep{Fagioli_2016ApJ...831..173F, van_der_Wel_2014ApJ...788...28V, Carnall_2023MNRAS.520.3974C}. Recent studies also suggest two distinct quenching pathways: (i) fast quenching associated with starbursts, which produces compact post-starburst galaxies, and (ii) slow quenching, which allows extended disks to persist, resulting in younger stellar populations \citep[e.g.,][]{Yano_2016ApJ...817L..21Y, Almaini_2017MNRAS.472.1401A, Maltby_2018MNRAS.480..381M, Wu_2018ApJ...868...37W, Suess_2020ApJ...899L..26S, Setton_2022ApJ...931...51S, Zhang_2024ApJ...976...36Z}. Hints of a connection between compactness with a recent burst of star formation has been reported for massive galaxies at even higher redshifts (e.g., $z \sim 1.5$ \citealt{Keating_2015ApJ...798...26K, Jain_2024OJAp....7E.113J}; $z\sim 2.5$--$3$ \citealt{Barro_2017ApJ...840...47B}). Despite these observations, the causality behind these empirical relations, as well as the basic link between SFHs, quenching mechanisms, and galaxy sizes remain difficult to establish \citep{Lilly_2016ApJ...833....1L}.

To address these questions, we adopted an 'archaeological' approach to trace the evolution of galaxies over time, examining how their SFHs vary with stellar mass and morphology. This study builds upon the work of \citet{Wu_2018ApJ...855...85W}, who analyzed the evolutionary pathways of a subsample of quiescent galaxies from the Large Early Galaxy Astrophysics Census \citep[LEGA-C;][]{van_der_Wel_2016ApJS..223...29V, van_der_Wel_2021ApJS..256...44V} DR2, using stellar age indicators such as \dfn~and \hd. Here, we investigate the relationship between galaxy sizes, SFHs, quenching times, and stellar population properties by leveraging spectrophotometric modeling results from {\tt Prospector} \citep{Leja_2017ApJ...837..170L, Johnson_2021ApJS..254...22J} and the LEGA-C DR3 dataset \citep{Nersesian_2025A&A...695A..86N}. By studying both the structural and stellar population properties of galaxies, we aim to determine whether there is a link between galaxy sizes, SFHs, and quenching times.

For this analysis, we used a sample of 2908 galaxies from the LEGA-C DR3 survey, which provides high signal-to-noise, high-resolution spectra of galaxies at redshifts $0.6 \leq z \leq 1$. These spectra enable precise measurements of stellar ages and SFHs \citep{Chauke_2018ApJ...861...13C, Wu_2018ApJ...855...85W, Barone_2022MNRAS.512.3828B, Kaushal_2024ApJ...961..118K, Nersesian_2024A&A...681A..94N, Nersesian_2025A&A...695A..86N, Gallazzi_2025arXiv251207952G, Gallazzi_2025arXiv251111805G}, while imaging from the \textit{Hubble Space Telescope} (HST) enables accurate size measurements. By linking galaxy structure, quenching mechanisms, and stellar population properties, this study contributes to a broader effort to understand the physical processes driving galaxy evolution.


\section{Data, sample, and analysis} \label{sec:data_and_sample}
   
\subsection{The LEGA-C survey} \label{subsec:legac}

LEGA-C is a $K_\mathrm{s}$-band magnitude selected spectroscopic survey, that obtained 4081 high S/N rest-frame optical spectra at a lookback time of $\sim 7$~Gyr \citep{van_der_Wel_2016ApJS..223...29V, van_der_Wel_2021ApJS..256...44V}. A LEGA-C spectrum has a typical observed wavelength range of 6300~\AA$~< \lambda <~$8800~\AA~or $\sim$ 3000~\AA~$< \lambda <~$5550~\AA~in restframe. The spectra belong to 3741 unique galaxy targets, with 340 spectra being duplicate observations that cover a slightly different spectral range. Each galaxy received $\sim 20$~hr of integration time at the 8~m Very Large Telescope (VLT), with an effective spectral resolution of $R\sim3500$. 

We refer the readers to \citet{van_der_Wel_2016ApJS..223...29V, van_der_Wel_2021ApJS..256...44V}, and \citet{Straatman_2018ApJS..239...27S}, for more details about the goals, design, and data reduction of the LEGA-C survey.

\subsection{SED fitting with {\tt Prospector}} \label{subsec:sed_fitting}

In \citet{Nersesian_2025A&A...695A..86N}, we derived the physical properties (i.e., stellar mass, SFR, stellar ages, and stellar metallicity) and SFHs of the LEGA-C galaxies, by fitting a subset of photometric bands from the UltraVISTA catalog \citep{Muzzin_2013ApJS..206....8M, Muzzin_2013ApJ...777...18M} together with the LEGA-C spectra. Only a subset of 8 broadbands from the UltraVISTA catalog was used ($B$, $V$, $r^+$, $i^+$, $z^+$, $Y$, $J$ and MIPS~$24~\mu$m). Then, the {\tt Prospector}\footnote{\url{https://github.com/bd-j/prospector}} inference framework \citep{Leja_2017ApJ...837..170L, Johnson_2021ApJS..254...22J} was applied to fit the broadband photometry and the LEGA-C spectroscopy, with a flexible spectrophotometric calibration. A physical model of 20 free parameters was generated that combined stellar, nebular, and dust models into composite stellar populations. The default SPS parameters in {\tt FSPS} were adopted (i.e., the {\tt MILES} stellar library, \citealt{Sanchez_Blazquez_2006MNRAS.371..703S}; and the {\tt MIST} isochrones, \citealt{Choi_2016ApJ...823..102C}), and a \citet{Chabrier_2003PASP..115..763C} initial mass function (IMF).

One of the particular advantages of {\tt Prospector} is the availability of nonparametric SFHs with various prior distributions and parameterizations. The benefit of using a nonparametric SFH is that we do not have to impose a particular function for the shape of the SFH. It has also been shown that a nonparametric SFH can outperform the traditional parametric SFH by vastly improving the stellar mass estimates \citep{Lower_2020ApJ...904...33L}. In this work we use the 'continuity' SFH with a Student's-t prior distribution \citep{Leja_2019ApJ...876....3L}. The 'continuity' prior favors a piecewise constant SFH with smooth transitions in SFR(t). To define the nonparametric SFH, we used eight time bins as free parameters, specified in lookback time. The first two time bins were fixed at 0--30~Myr and 30--100~Myr to capture any recent variations in the SFH. For the full description of the SED modeling with {\tt Prospector} and a presentation of the various quality tests of the fits, we refer the readers to \citet{Nersesian_2025A&A...695A..86N}. 

\begin{figure}[t]
    \centering
    \includegraphics[width=\columnwidth]{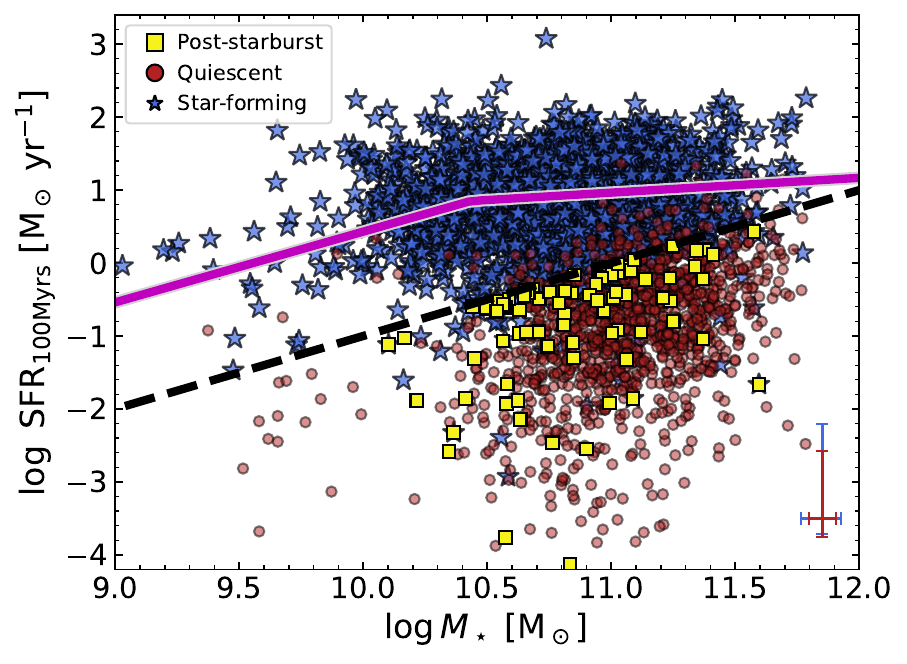}
    \caption{\sfr~as a function of $M_\star$ for the LEGA-C galaxies at $0.6 \le z_\mathrm{spec} \le 1$. Galaxies are color-coded by their $UVJ$ diagram classification as $UVJ$ star-forming (blue stars) and $UVJ$ quiescent (red points). Post-starburst galaxies are indicated as yellow squares. The magenta line indicates the SFS at $z = 0.6$ from \citet{Leja_2022ApJ...936..165L}, and the dashed black line indicates the separation between star formation and quiescence (sSFR = $10^{-11}$~yr$^{-1}$). The average uncertainties for the two main galaxy populations are shown in the bottom-right corner in their respective colors.}
    \label{fig:fig_1}
\end{figure}

\subsection{Primary galaxy sample} \label{subsec:final_sample}

In this paper, we used the same sample of 2908 galaxies presented in \citet{Nersesian_2025A&A...695A..86N}. The sample was selected to be all galaxies that have a {\tt PRIMARY} flag equal to one, indicating that a galaxy was chosen from the $K_\mathrm{s}$-band selected parent sample \citep{van_der_Wel_2021ApJS..256...44V}. The redshift range of our sample is $0.6 \le z \le 1$, with a median redshift $\langle z \rangle = 0.76^{+0.18}_{-0.09}$, and an average S/N of $\sim17.2$~\AA$^{-1}$ for the continuum. This sample is characterized by a median stellar velocity dispersion $\langle \sigma_\star \rangle = 166^{+40.1}_{-39.3}~\mathrm{km}~\mathrm{s}^{-1}$, a median stellar mass of $\log \langle M_\star / \rm{M}_\odot \rangle = 10.93^{+0.23}_{-0.28}$, and a median SFR over the last 100~Myr $\log~\langle$\sfr~/$\rm{M}_{\odot}~\rm{yr}^{-1} \rangle = 0.25^{+0.66}_{-0.94}$. The subscripts/superscripts indicate the $16^\mathrm{th}$--$84^\mathrm{th}$ percentile ranges.

The completeness of the LEGA-C spectroscopic sample is well understood thanks to its $K_\mathrm{s}$-band selection and high measurement success rate \citep[see Appendix~A of][]{van_der_Wel_2021ApJS..256...44V}. This translates into a representative sampling of the $0.6 < z < 1.0$ galaxy population, for which the completeness depends only on redshift and $K_\mathrm{s}$-band magnitude, and no other galaxy property.

Figure~\ref{fig:fig_1} shows the relationship between \sfr~and $M_\star$, otherwise known as the star-forming sequence \citep[SFS;][]{Strateva_2001AJ....122.1861S, Baldry_2004ApJ...600..681B, Daddi_2007ApJ...670..156D, Noeske_2007ApJ...660L..43N, Elbaz_2007A&A...468...33E, Whitaker_2012ApJ...754L..29W}. Galaxies can be separated into two broad galaxy populations, quiescent and star-forming. It has been shown that the normalization of the SFS evolves with redshift \citep[e.g.,][]{Whitaker_2012ApJ...754L..29W, Fumagalli_2014ApJ...796...35F, Speagle_2014ApJS..214...15S, Daddi_2022A&A...661L...7D, Leja_2022ApJ...936..165L, Popesso_2023MNRAS.519.1526P, Koprowski_2024A&A...691A.164K}. In this work, we used the definition of SFS ridge at $z=0.6$ from \citet{Leja_2022ApJ...936..165L}. We find a rather flat relation between SFR and stellar mass for the star-forming galaxies. A number of recent studies have also observed the flattening of the SFS for massive galaxies ($M_\star > 10^{10}~$M$_\odot$) \citep[e.g.,][]{Lee_2015ApJ...801...80L, Schreiber_2016A&A...589A..35S, Leslie_2020ApJ...899...58L, Daddi_2022A&A...661L...7D, Leja_2022ApJ...936..165L, Popesso_2023MNRAS.519.1526P}. However, the cause remains debated, with proposed explanations including reduced star-formation efficiency \citep[e.g.,][]{Schreiber_2016A&A...589A..35S}, a decline in available cold molecular gas \citep[e.g.,][]{Daddi_2022A&A...661L...7D, Popesso_2023MNRAS.519.1526P}, or the result of mergers rather than ongoing star formation \citep[e.g.,][]{Fu_2024MNRAS.532..177F}. Quiescent galaxies show a stronger correlation with stellar mass, forming their own 'red' sequence. At fixed mass, quiescent galaxies at intermediate redshift are characterized by an older stellar population, high metallicity, and a more concentrated light profile than star-forming galaxies \citep[e.g.,][]{Beverage_2021ApJ...917L...1B, Cappellari_2023MNRAS.526.3273C, Martorano_2025A&A...694A..76M, Nersesian_2025A&A...695A..86N, Gallazzi_2025arXiv251111805G}.

Different criteria are usually employed to classify galaxies as quiescent. The two most common criteria to separate galaxies are based on (i) their locus on the $UVJ$ diagram \citep{Labbe_2005ApJ...624L..81L, Wuyts_2007ApJ...655...51W, Williams_2009ApJ...691.1879W}, and (ii) their specific star-formation rate ($\mathrm{sSFR} = \mathrm{SFR}_\mathrm{100Myrs}/M_\star$). Using the definition of \citet{Muzzin_2013ApJ...777...18M} for quiescence in the $UVJ$ diagram, we find 1208 quiescent and 1700 star-forming galaxies. Alternatively, quiescence can be defined based on the sSFR of galaxies. The sSFR can be seen as a measure of the current to past star formation in galaxies, tracing the hardness of the UV radiation field \citep{Ciesla_2014A&A...565A.128C}. Many studies suggest a threshold at $\log \mathrm{sSFR} / \mathrm{yr}^{-1} \approx -10$ \citep[e.g.,][]{Whitaker_2017ApJ...838...19W, Wu_2018ApJ...868...37W, Leja_2019ApJ...880L...9L}, while others propose a slightly lower boundary at $\log \mathrm{sSFR} / \mathrm{yr}^{-1} \approx -10.55$ \citep[e.g.,][]{Gallazzi_2014ApJ...788...72G}. In this study, we adopt a threshold of $\log \mathrm{sSFR} / \mathrm{yr}^{-1} \le -11$ \citep[i.e.,][]{Brinchmann_2004MNRAS.351.1151B, Fontanot_2009MNRAS.397.1776F, Cecchi_2019ApJ...880L..14C, Donnari_2019MNRAS.485.4817D, Paspaliaris_2023A&A...669A..11P, Nersesian_2025A&A...695A..86N}, which aligns best with the $UVJ$ quiescence criteria, yielding 1240 quiescent and 1668 star-forming galaxies in our sample.

In Fig.~\ref{fig:fig_1}, galaxies are color-coded by their $UVJ$ diagram classification as star-forming (blue stars) and quiescent (red points). It is clear from Fig.~\ref{fig:fig_1} that the $UVJ$ criterion for quiescence largely follows the sSFR criterion, albeit with 7\% of $UVJ$-quiescent galaxies showing signs of star formation ($\mathrm{sSFR} \ge 10^{-11}$~yr$^{-1}$). Furthermore, a $UVJ$ quiescent selection can yield a 10\%--30\% contamination from star-forming galaxies \citep{Moresco_2013A&A...558A..61M, Belli_2017ApJ...841L...6B, Fang_2018ApJ...858..100F, Merlin_2018MNRAS.473.2098M, Schreiber_2018A&A...618A..85S, Leja_2019ApJ...880L...9L}. These contamination fractions are indicative of the slight challenge of defining quiescence. When necessary, we divided the sample into star-forming and quiescent galaxies based on the sSFR criterion as a more robust method to determine quiescence. The results and conclusions of this paper are not affected by the definition of quiescence, since the demographics we obtain from the two quiescence criteria are quite similar. 

\subsection{Post-starburst galaxies} \label{subsec:psb}

Post-starburst galaxies (PSBs) are interesting objects to study as they can hint at the different pathways that galaxies follow towards quiescence \citep[e.g.,][]{Wu_2018ApJ...868...37W}. The spectrum of a PSB galaxy is characterized by prominent Balmer absorption lines, indicative of an abundance of A-type stars \citep{Dressler_1983ApJ...270....7D, Balogh_1999ApJ...527...54B,  Dressler_1999ApJS..122...51D}, and very weak nebular emission, indicative of the absence of O- and B-type stars. The spectroscopic information of PSBs suggest a quenching event that occurred in a relatively short timescale (typically less than $1$~Gyr) after a burst of star-formation.

Similar to the studies of \citet{Wu_2018ApJ...868...37W, Wu_2020ApJ...888...77W}, we selected the PSBs in the quiescent galaxy population based upon their \hda~absorption strength \citep{Gallazzi_2025arXiv251207952G}. In particular, we required quiescent galaxies to have a \hda~$ \ge 4$~\AA. If we apply this criterion we end up with 84 PSB galaxies in our sample, with a median value of \hda~$= 5.42$~\AA. PSBs are marked as yellow squares in Fig.~\ref{fig:fig_1}. The fraction of PSBs among the quiescent galaxy population is $\sim7\%$, and is in agreement with PSB fractions found in previous studies \citep[e.g.,][]{Poggianti_2009ApJ...693..112P, Gallazzi_2014ApJ...788...72G, Wu_2020ApJ...888...77W}. It is important to note that the PSB sample reflects a selection based on spectral features, rather than necessarily representing a physically distinct class of galaxies.

\begin{figure}[t]
    \centering
    \includegraphics[width=\columnwidth]{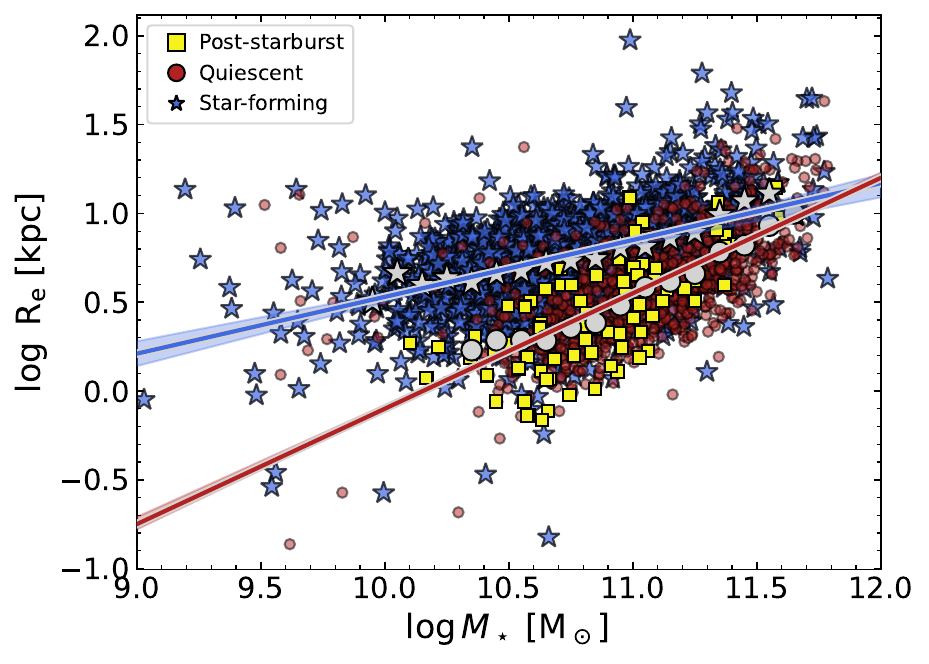}
    \caption{Galaxy size as a function of $M_\star$ for the LEGA-C galaxies at $0.6 \le z_\mathrm{spec} \le 1$. Based on their sSFR, galaxies are separated into star-forming (blue stars) and quiescent (red points) populations. Post-starburst galaxies are indicated as yellow squares. For each galaxy population, we fitted a linear relation between the median sizes in each 0.1~dex mass bin and the stellar mass (gray symbols). To avoid the flattening of the relations at the low-mass end, we fit star-forming galaxies with $\log \left(M_\star / \mathrm{M}_\odot \right) \ge 9.8$, and quiescent galaxies with $\log \left(M_\star / \mathrm{M}_\odot \right) \ge 10.3$.}
    \label{fig:fig_2}
\end{figure}

\subsection{The Size--Mass relation} \label{subsec:mass_size_relation}

Figure~\ref{fig:fig_2} shows the relation between galaxy size and stellar mass. \citet{Wu_2018ApJ...868...37W} measured the galaxy sizes by fitting a 2D \citet{Sersic_1963BAAA....6...41S} profile to the \textit{HST} ACS F814W images from COSMOS \citep{Scoville_2007ApJS..172....1S}. In particular, a $10\arcsec$ cutout for each galaxy was created and then fitted with {\tt galfit} \citep{Peng_2010ApJ...721..193P}, using a single S{\'e}rsic profile with 6 free parameters (S{\'e}rsic index, effective radius, total magnitude, axis ratio, position angle, and background). The effective radius ($R_\mathrm{e}$) is the semimajor axis of the ellipse that contains half of the total flux of the best-fit S{\'e}rsic model. Out of the 2908 galaxies, only 2744 have a size measurement, of which 1577 are star-forming galaxies and 1167 quiescent.  

To derive the size--mass relation for our LEGA-C primary sample, we followed the approach of \citet{Wu_2018ApJ...868...37W} and fitted a linear relation between the median sizes and the stellar mass in each 0.1~dex mass bin, in log--log space. We restrict the range of mass that is fitted to the high mass regime, in order to exclude the low-mass range where the mass-size relations are known to flatten \citep[e.g.,][]{Graham_2013pss6.book...91G, Cappellari_2013MNRAS.432.1862C, Shankar_2014MNRAS.439.3189S, Norris_2014MNRAS.443.1151N, van_der_Wel_2014ApJ...788...28V, Lange_2015MNRAS.447.2603L, Furlong_2017MNRAS.465..722F, Whitaker_2017ApJ...838...19W, Martorano_2024ApJ...972..134M}. The following functional form was used:

\begin{equation} \label{eq:size_mass_relation}
\\\\\ \log \left(R_\mathrm{e} / \mathrm{kpc} \right)_\mathrm{med} = \alpha \times \left[\log \left(M_\star / \mathrm{M}_\odot\right) - 11 \right] + \beta,
\end{equation}

\noindent where ($\alpha$, $\beta$) are the best-fit slope and intercept respectively. We estimated the median and 1--$\sigma$ range of these two free parameters using a bootstrap resampling (1000 iterations), considering the uncertainties on $M_\star$. For the star-forming galaxies, we find a relation with a median slope of $\alpha = 0.32 \pm 0.04$ and a median intercept of $\beta = 0.84 \pm 0.02$. For the quiescent galaxies, we measure a much steeper slope with $\alpha = 0.65 \pm 0.02$ and $\beta = 0.55 \pm 0.01$. Our measured slopes are in good agreement with those derived for late- and early-type galaxies at comparable redshift ($z \simeq 0.75$) in the CANDELS survey \citep[$0.22 \pm 0.01$, and $0.71 \pm 0.01$ respectively][]{van_der_Wel_2014ApJ...788...28V}. The best-fit relations are shown in Fig.~\ref{fig:fig_2} with the blue (star-forming galaxies) and red (quiescent galaxies) lines.

For completeness, we also fit the size--mass relation for the full sample. Again, to avoid the flattening of the relation at the low-mass end, we fit only those galaxies with $\log \left(M_\star / \mathrm{M}_\odot \right) \ge 10.3$. We find a relation with a median slope of $\alpha = 0.31 \pm 0.01$ and a median intercept of $\beta = 0.73 \pm 0.01$. All three best-fit lines of the size--mass relation are summarized in Table~\ref{tab:table_1}.  

\begin{figure}[t]
    \centering
    \includegraphics[width=\columnwidth]{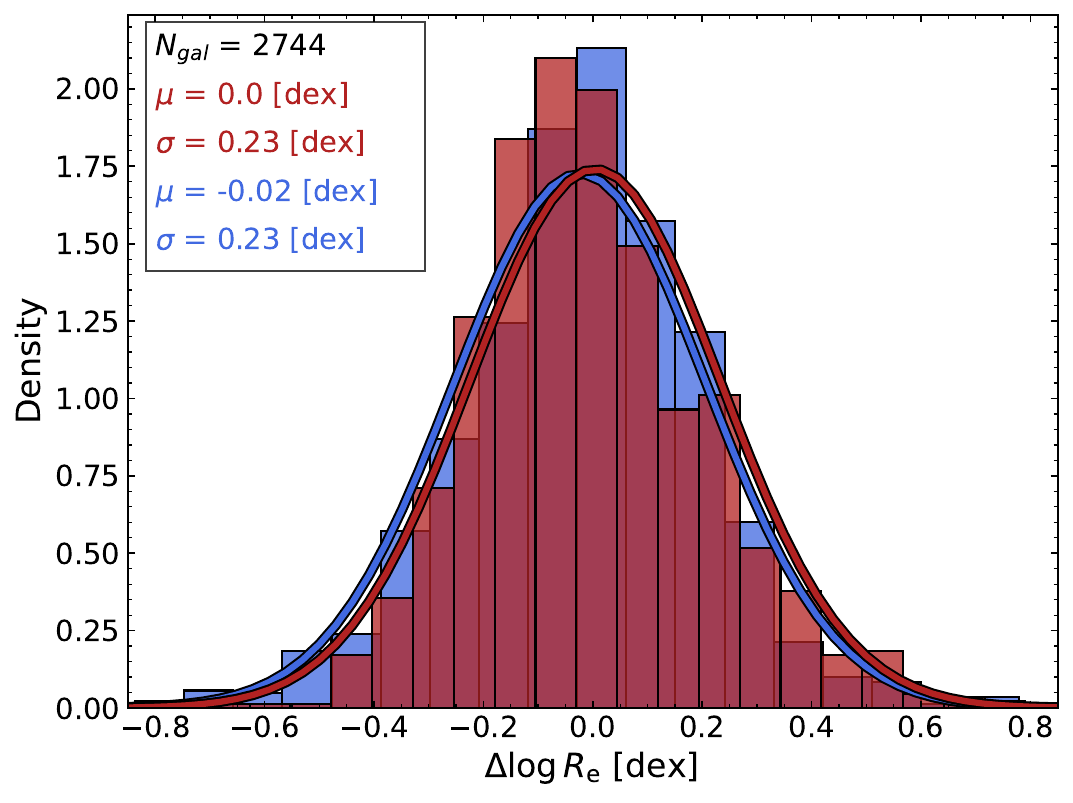}
    \caption{Distribution of the galaxy size metric $\Delta \log \left(R_{e}\right)$, defined as the distance of the sizes $R_\mathrm{e}$ from the median size--mass relation at a given mass, for each galaxy population. The blue histogram shows the relative size distribution for the star-forming galaxies, while the red histogram shows the relative size distribution for the quiescent ones. The corresponding solid curves are the best-fit Gaussian distributions. The statistics of the mean offset ($\mu$) and variance ($\sigma$) of the each Gaussian distribution is shown at the top-left corner.}
    \label{fig:fig_3}
\end{figure}

\begin{table}[t]
    \caption{Best-fit lines of the size--mass relation as given in Eq.~\ref{eq:size_mass_relation}.}
    \centering
    \scalebox{0.9}{
    \begin{tabular}{l c c}
    \hline
    \hline
    Sample & Slope ($\alpha$) & Intercept at $10^{11}~\mathrm{M}_\odot$ ($\beta$) \\ 
    \hline
    All galaxies & $0.31\pm0.01$ & $0.73\pm0.01$ \\
    Star-forming galaxies & $0.32\pm0.04$ & $0.84\pm0.02$ \\
    Quiescent galaxies & $0.65\pm0.02$ & $0.55\pm0.01$ \\
    \hline
    \end{tabular}}
    \tablefoot{We provide the best-fit slopes and intercepts of the size--mass relation for the full sample, star-forming, and quiescent galaxies. We classify galaxies as quiescent based on the sSFR criterion described in Sect.~\ref{subsec:final_sample}.}
    \label{tab:table_1}
\end{table}

\begin{figure*}[t]
    \centering
    \includegraphics[width=\textwidth]{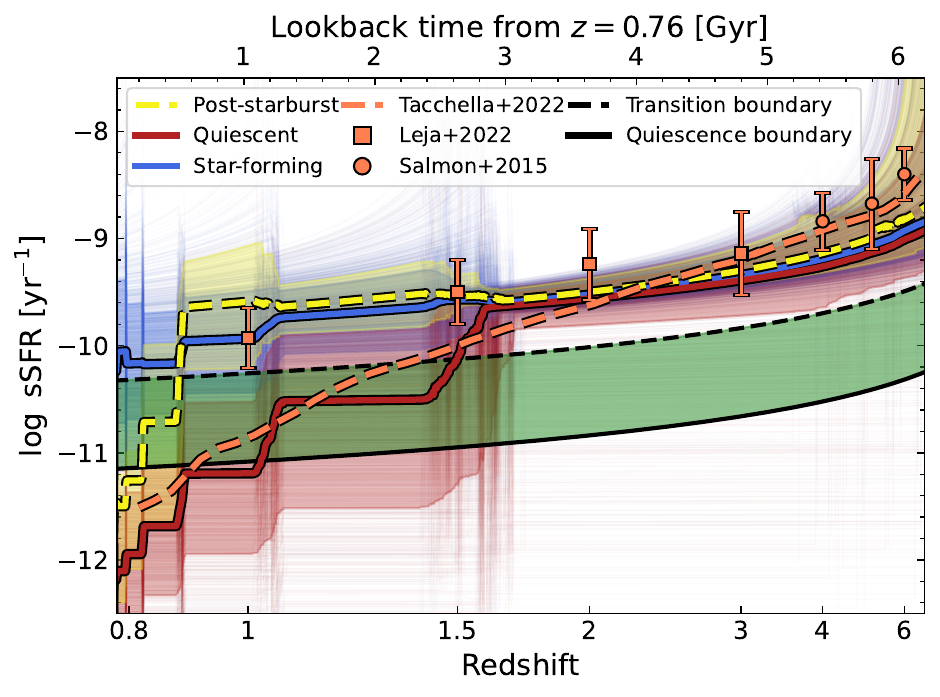}
    \caption{Nonparametric SFHs of LEGA-C galaxies recovered with {\tt Prospector}. The fitted SFR of each time bin is normalized by the total stellar mass formed up to that time bin. The SFHs are divided into the star-forming (blue) and quiescent (red) populations, based on their sSFR. In the background, we show the individual SFHs of our sample, while the thick lines and shaded areas denote the median and the $16^\mathrm{th}$--$84^\mathrm{th}$ percentile range for the star forming (solid blue line and shaded blue area), quiescent (solid red line and shaded red area; excluding the 84 PSBs), and PSB (dashed yellow line and shaded yellow area) galaxies. The green shaded area displays the transition boundary from star forming (dashed black line) to quiescence (solid black line). The coral point and square markers show the measurements of sSFR by \citet{Salmon_2015ApJ...799..183S} and \citet{Leja_2022ApJ...936..165L} respectively, at higher redshift. The dashed coral line shows the SFH of lower-mass, quiescent galaxies ($10.5 \leq \log (M_\star / {\rm M_\odot}) < 11$) from \citet{Tacchella_2022ApJ...926..134T}.}
    \label{fig:fig_4}
\end{figure*}

To quantify the galaxy sizes within each population, we calculated the distance of the sizes $R_\mathrm{e}$ from the median size--mass relation of its respective population at a given stellar mass, according to Equation~\ref{eq:size_mass_relation}. This was computed using the following function:

\begin{equation} \label{eq:delta_size}
\\\\\ \Delta \log \left(R_{e}\right) = \log \left(R_\mathrm{e}/\mathrm{kpc}\right) - \log \left(R_\mathrm{e}/\mathrm{kpc}\right)_\mathrm{med}.
\end{equation}

\noindent Figure~\ref{fig:fig_3} shows the distributions of $\Delta \log \left(R_{e}\right)$ for both the star-forming and quiescent galaxies. Both distributions are centered very close to zero with a small offset of $\mu = -0.02$~dex for the star-forming galaxies. We also measure a similar standard deviation ($\sigma$) of 0.23~dex for both galaxy populations. The standard deviation in this work is consistent with the observed scatter expected at an average redshift of 0.75 \citep[see Fig.~6 from][]{van_der_Wel_2014ApJ...788...28V}. 

\begin{figure*}[t]
    \centering
    \includegraphics[width=\textwidth]{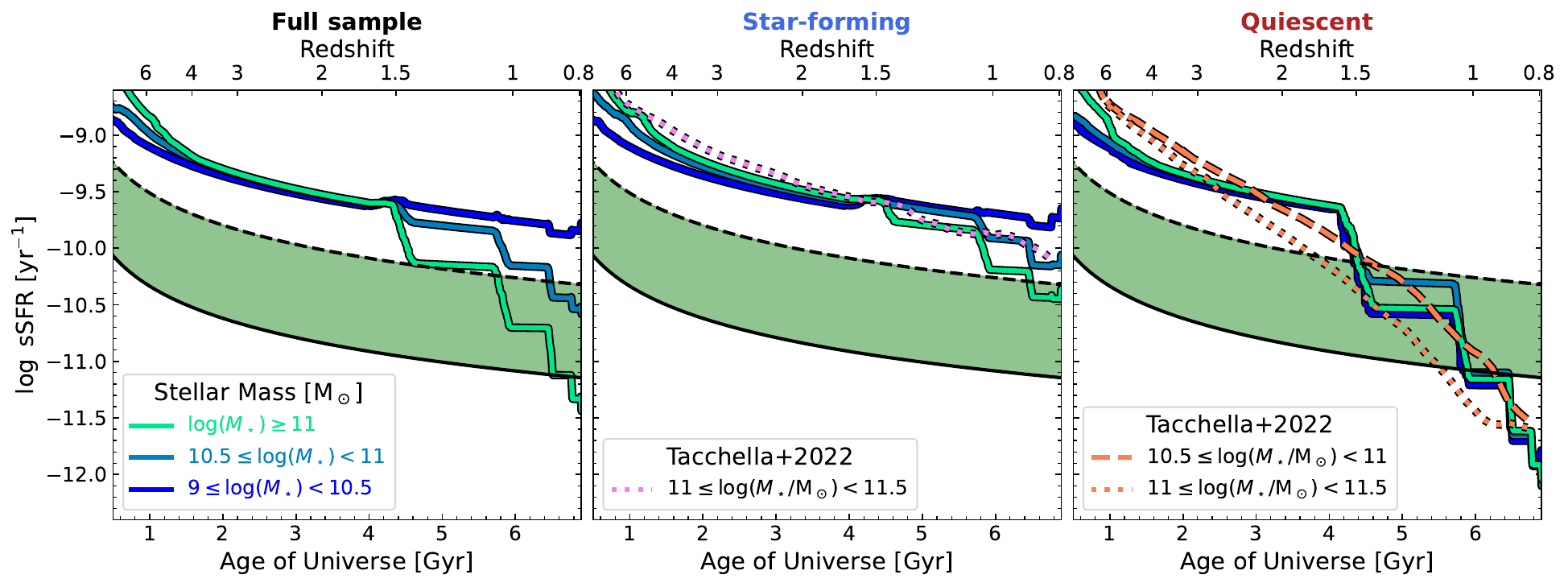}
    \caption{Median SFHs of LEGA-C galaxies separated into three stellar mass bins. From left to right: median SFHs of the full primary galaxy sample, the star-forming population, and the quiescent population. The dotted violet line in the star-forming galaxies' panel shows the SFH of high-mass, star-forming galaxies from \citet{Tacchella_2022ApJ...926..134T}. The dashed and dotted coral lines in the quiescent galaxies' panel show the SFHs of lower-mass and higher-mass, quiescent galaxies from \citet{Tacchella_2022ApJ...926..134T}, respectively. The green shaded area displays the transition boundary from star forming (dashed black line) to quiescence (solid black line).}
    \label{fig:fig_5}
\end{figure*}

\section{The SFHs of the LEGA-C galaxy sample} \label{sec:sfhs}

The SFHs of LEGA-C galaxies is a topic that has been explored in the past. A first reconstruction of SFHs for a small subsample of DR2 LEGA-C galaxies was performed by \citet{Chauke_2018ApJ...861...13C}. A more comprehensive study on the SFHs of DR3 LEGA-C galaxies was recently performed by \citet{Kaushal_2024ApJ...961..118K}. The authors conducted a spectrophotometric analysis using {\tt BAGPIPES} and assuming a parametric double-power law SFH. Their goal was to quantify the timescales of stellar mass formation. \citet{Kaushal_2024ApJ...961..118K} also performed a comparison between their SFHs and the nonparametric SFHs recovered with {\tt Prospector} \citep{Nersesian_2025A&A...695A..86N}. Their analysis yielded consistent late-time SFRs for star-forming and quiescent galaxies, despite the differences in the modeling methods (parametric versus nonparametric). Here, we give a brief overview of the nonparametric SFHs of the LEGA-C survey as retrieved with {\tt Prospector} for the primary galaxy sample (2908 galaxies), and compare our results with other observations at higher $z_\mathrm{obs}$. For a more detailed analysis and discussion on the SFHs of the LEGA-C sample and the formation times, we refer to \citet[][see also recent study by \citealt{Wan_2025MNRAS.539.2891W}]{Kaushal_2024ApJ...961..118K}. 

\subsection{The variety of SFHs} \label{subsec:sfh_variety}
 
We defined the SFHs using the sSFR($t$) instead of SFR($t$), because the sSFR is a better indicator of quiescence \citep{Whitaker_2017ApJ...838...19W}, with its inverse tracing the formation timescale of a galaxy's stellar population. The sSFR($t$) is simply defined as the fitted SFR of each time bin in our physical model, normalized by the total stellar mass formed up to that time bin ($\mathrm{sSFR}(t) = \mathrm{SFR}(t)/M_\mathrm{formed}$). Then, we interpolated all SFHs in our primary sample on a uniform age grid from 0.01--7~Gyr, and smoothed using a {\tt Box1DKernel} with a smoothing factor of 10~Myr \citep[see][]{Kaushal_2024ApJ...961..118K}. In Fig.~\ref{fig:fig_4}, we plot the individual SFHs of our primary sample, as well as the median trends for the three galaxy populations: the star forming (solid blue line), the quiescent (solid red line), and the PSBs (dashed yellow line).\footnote{In fact, \citet{Suess_2022ApJ...935..146S} demonstrated that {\tt Prospector} provides a very accurate reconstruction of PSB SFHs.} We also plot the transition boundary (green region) from star forming (${\rm sSFR} = 1 / \left[3~t_{\rm H}(z)\right]$) to quiescence (sSFR$ = 1/\left[20~t_{\rm H}(z)\right]$), as defined by \citet{Tacchella_2022ApJ...926..134T} \citep[see also][]{Pacifici_2016ApJ...832...79P}, where $t_{\rm H}(z)$ is the Hubble time at a median redshift $z = 0.76$. We define the quenching timescale ($\tau_\mathrm{q}$) as the time a galaxy takes to transition from star forming to quiescent. 

Looking at the individual SFHs, the LEGA-C galaxies have a large range of SFHs, spanning five orders of magnitude in sSFR. The $\tau_\mathrm{q}$ depends on the precise definition of the boundary region; we address this caveat in Sect.~\ref{sec:discussion} and Appendix~\ref{apdx:A}. Moreover, the apparent sharp changes in sSFR at specific lookback times (e.g., 0.5, 1.1, 2.8~Gyr) reflect artifacts introduced by the binning scheme rather than physical events across the galaxy population. While the overall trends in quenching timescales are robust, the precise values of $\tau_q$ should be interpreted with caution.

The large diversity of pathways towards quiescence is evident from the $16^\mathrm{th}$--$84^\mathrm{th}$ percentile range of the red galaxies. Looking at the individual SFHs of red galaxies, we notice that galaxies can reach quiescence quite early (within the first 2~Gyr after the Big Bang) or at a later epoch, with different quenching timescales. 

The median trends in Fig.~\ref{fig:fig_4} suggest that within the first 4~Gyr after the Big Bang the reconstructed SFHs of quiescent and star-forming galaxies are broadly similar. However, the constraints at high $z$ ($z>3$) are weak and largely shaped by the modeling assumptions rather than direct information in the data. In this regime, our median SFHs fall systematically below the photometry-based sSFR measurements of \citet{Salmon_2015ApJ...799..183S}. This qualitative comparison should not be interpreted as a direct tension between equally robust measurements, as the underlying methods and data differ. Nevertheless, our results are still consistent with \citet{Salmon_2015ApJ...799..183S} within the $16^\mathrm{th}$--$84^\mathrm{th}$ percentile range. At lower redshift ($z<3$), the median SFH of the star-forming galaxies (solid blue line) agrees very well with the sSFR measurements of \citet{Leja_2022ApJ...936..165L}. Likewise, the median SFH of the quiescent population (solid red line) closely follows that of lower-mass quiescent galaxies ($10.5 \leq \log (M_\star / {\rm M_\odot}) < 11$) from \citet{Tacchella_2022ApJ...926..134T}. Based on these median SFH trends, we infer that quiescent galaxies at $0.6 \leq z \leq 1$ began to diverge from the star-forming population around $z \approx 1.5$, entering a transition phase toward quiescence with a broad range of quenching timescales. In contrast, the median SFH of PSB galaxies shows a distinctly different pattern, maintaining high SFRs until undergoing a sudden and rapid quenching episode. We emphasize, however, that this behavior is partly a consequence of our spectroscopic PSB selection, which preferentially identifies galaxies with recent, rapid transitions. Similar fast-quenching events may also have occurred in galaxies now classified as quiescent, but they are more difficult to identify at earlier epochs.

From the individual SFHs of the quiescent population, we calculated the median quenching timescale obtaining $\langle\tau_\mathrm{q}\rangle = 1.23\pm0.04$~Gyr ($16^\mathrm{th}$--$84^\mathrm{th}$ percentile range: 0.33--2.4~Gyr) for the quiescent galaxies, and $\langle\tau_\mathrm{q}\rangle = 0.13 \pm 0.03$~Gyr ($16^\mathrm{th}$--$84^\mathrm{th}$ percentile range: 0.01--0.47~Gyr) for the PSBs. The uncertainties represent the statistical uncertainty on the median ($\sigma/\sqrt{N}$), while the $16^\mathrm{th}$--$84^\mathrm{th}$ range reflects the intrinsic spread among galaxies. Notably, many quiescent galaxies also exhibit relatively short $\tau_\mathrm{q}$ indicating a diversity of quenching pathways.

\begin{figure*}[t]
    \centering
    \includegraphics[width=\textwidth]{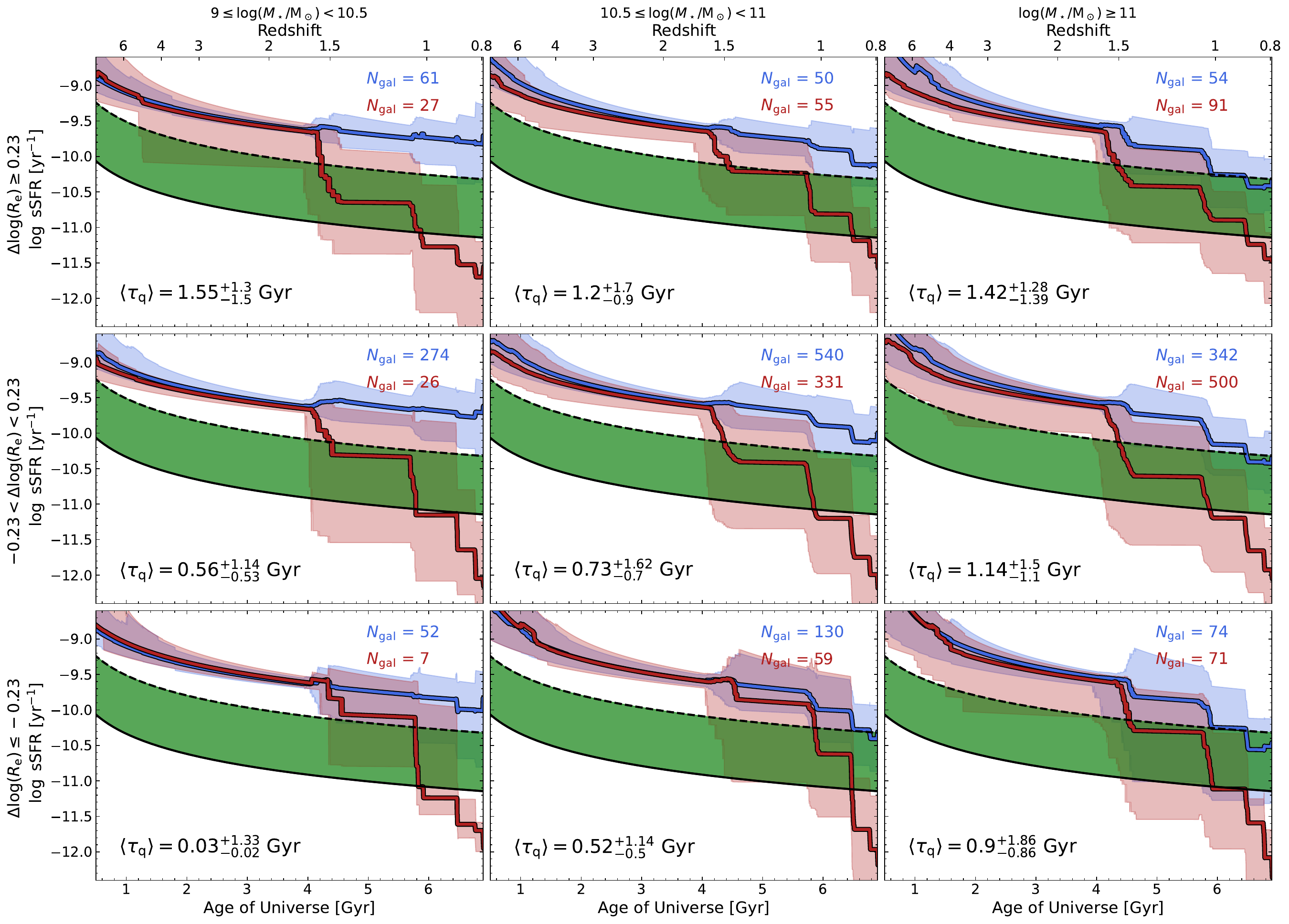}
    \caption{Median SFHs of LEGA-C galaxies separated into three stellar mass and three galaxy size bins: $\Delta \log (R_\mathrm{e}) \leq -0.23$, $-0.23 < \Delta \log (R_\mathrm{e}) < 0.23$, and $\Delta \log (R_\mathrm{e}) \geq 0.23$. The SFHs are divided into the star-forming (blue) and quiescent (red) populations, based on their sSFR. The corresponding shaded regions denote the $16^\mathrm{th}$--$84^\mathrm{th}$ percentile range. The green shaded area displays the transition boundary from star forming (dashed black line) to quiescence (solid black line). In the bottom-left corner of each panel, we report the median $\langle\tau_\mathrm{q}\rangle$ from the individual SFHs in each bin (shown in black), along with the corresponding $16^\mathrm{th}$--$84^\mathrm{th}$ percentile range.}
    \label{fig:fig_6}
\end{figure*}

\subsection{SFHs as a function of $M_\star$} \label{subsec:sfh_mstar}

In order to assess whether galaxies of different mass follow different evolutionary trajectories, we binned their SFHs into three stellar mass bins, and calculated the median SFH in each bin. In Fig.~\ref{fig:fig_5}, we present the median SFHs as a function of stellar mass for the full sample, star-forming, and quiescent galaxies. The median SFHs of the full sample indicate that galaxies with varying stellar masses exhibit distinct sSFRs from an early stage (within the first $\sim 1.5$~Gyr after the Big Bang). In particular, higher-mass galaxies seem to form stars faster than the less massive galaxies. From $\sim$ 2 to 4~Gyr after the Big Bang, we see a decline in star formation with time, with all galaxies having very similar sSFRs. At $z \sim 1.5$, we observe that the most massive galaxies begin to transition from star formation to quiescence. Our results support the downsizing scenario, stating that the highest SFR occurs in progressively less massive galaxies with decreasing redshift \citep{Cowie_2008ApJ...686...72C}, and thus more massive galaxies contain older stellar populations, consistent with previous studies on this topic \citep[e.g.][]{Fontanot_2009MNRAS.397.1776F, Tojeiro_2009ApJS..185....1T, McDermid_2015MNRAS.448.3484M, Ibarra_Medel_2016MNRAS.463.2799I, Pacifici_2016ApJ...832...79P, Chauke_2018ApJ...861...13C, Tacchella_2022ApJ...926..134T, Kaushal_2024ApJ...961..118K, Gallazzi_2005MNRAS.362...41G, Gallazzi_2014ApJ...788...72G}.

Analyzing the SFHs of the star-forming galaxies reveals that more massive galaxies began transitioning toward quiescence at $z \sim 1$ and remain in this phase, while less massive galaxies continue forming stars, albeit at a slower rate. Additionally, the median SFHs for galaxies in the intermediate stellar mass bin ($10.5 \leq \log (M_\star / {\rm M_\odot}) < 11$) align perfectly with the median SFH reported for the same mass bin in \citet{Tacchella_2022ApJ...926..134T}. Quiescent galaxies display a weaker dependence on $M_\star$, with the transition to quiescence occurring at approximately the same epoch ($z\sim 1.5$). Based on the median SFHs, we find a weak correlation between stellar mass and quenching time, with an average quiescence transition duration of $\tau_\mathrm{q} \approx 1.2$~Gyr. These findings are consistent with the quiescence times reported by both \citet{Kaushal_2024ApJ...961..118K} and \citet{Tacchella_2022ApJ...926..134T}. This agreement is particularly reassuring, as \citet{Kaushal_2024ApJ...961..118K} analyzed the same LEGA-C dataset but applied a different fitting method (i.e., {\tt BAGPIPES}), whereas \citet{Tacchella_2022ApJ...926..134T} utilized the same fitting method (i.e., {\tt Prospector}) but analyzed a different dataset \citep[HALO7D program, a survey conducted in CANDELS fields][]{Grogin_2011ApJS..197...35G, Koekemoer_2011ApJS..197...36K}.  

\subsection{Dependence of SFHs on galaxy size} \label{subsec:sfh_size_mass}

To explore the connection between galaxy size and SFHs, we estimated the median SFHs of LEGA-C galaxies, dividing them into three stellar mass bins and three size bins, as parameterized by the relative size metric $\Delta \log (R_\mathrm{e})$. Specifically, $\Delta \log (R_\mathrm{e})$ for each galaxy population was determined based on their corresponding size--mass relation (see Table~\ref{tab:table_1} for definitions). The solid lines in Fig.~\ref{fig:fig_6} represent the median SFH trends for star-forming (blue) and quiescent (red; including the 84 PSBs) galaxies, while the corresponding shaded regions denote the $16^\mathrm{th}$--$84^\mathrm{th}$ percentile range. 

Examining the median SFHs of star-forming galaxies, we find that at a fixed size, the most massive galaxies cross the transition boundary (dashed black line) earlier than lower-mass galaxies. This observation aligns with the downsizing scenario, where more massive galaxies tend to quench earlier. At a fixed mass, we do not find strong evidence of a correlation between $\Delta \log (R_\mathrm{e})$ and the average sSFR computed over the last 100~Myr. That said, there is a very subtle indication that at fixed mass larger galaxies ($\Delta \log (R_\mathrm{e}) > 0.23$~dex) sustain star-formation for a longer duration compared to their smaller counterparts ($\Delta \log (R_\mathrm{e}) \leq -0.23$~dex).

In contrast, for quiescent galaxies at a fixed mass, we observe a positive correlation between quenching timescale and $\Delta \log (R_\mathrm{e})$. Specifically, larger quiescent galaxies exhibit longer quenching timescales than smaller ones. For example, among intermediate-mass quiescent galaxies, the median quenching timescale for galaxies $1\sigma$ below the size--mass relation is $\langle\tau_\mathrm{q}\rangle = 0.52$~Gyr, whereas larger galaxies in the same mass range quench approximately 0.7~Gyr more slowly ($\langle\tau_\mathrm{q}\rangle = 1.2$~Gyr). At a fixed size, we also find tentative evidence that more massive galaxies quench at a slower rate.

\section{Galaxy quenching and the size--mass plane} \label{sec:tq_size_mass}

\begin{figure}[t]
    \centering
    \includegraphics[width=\columnwidth]{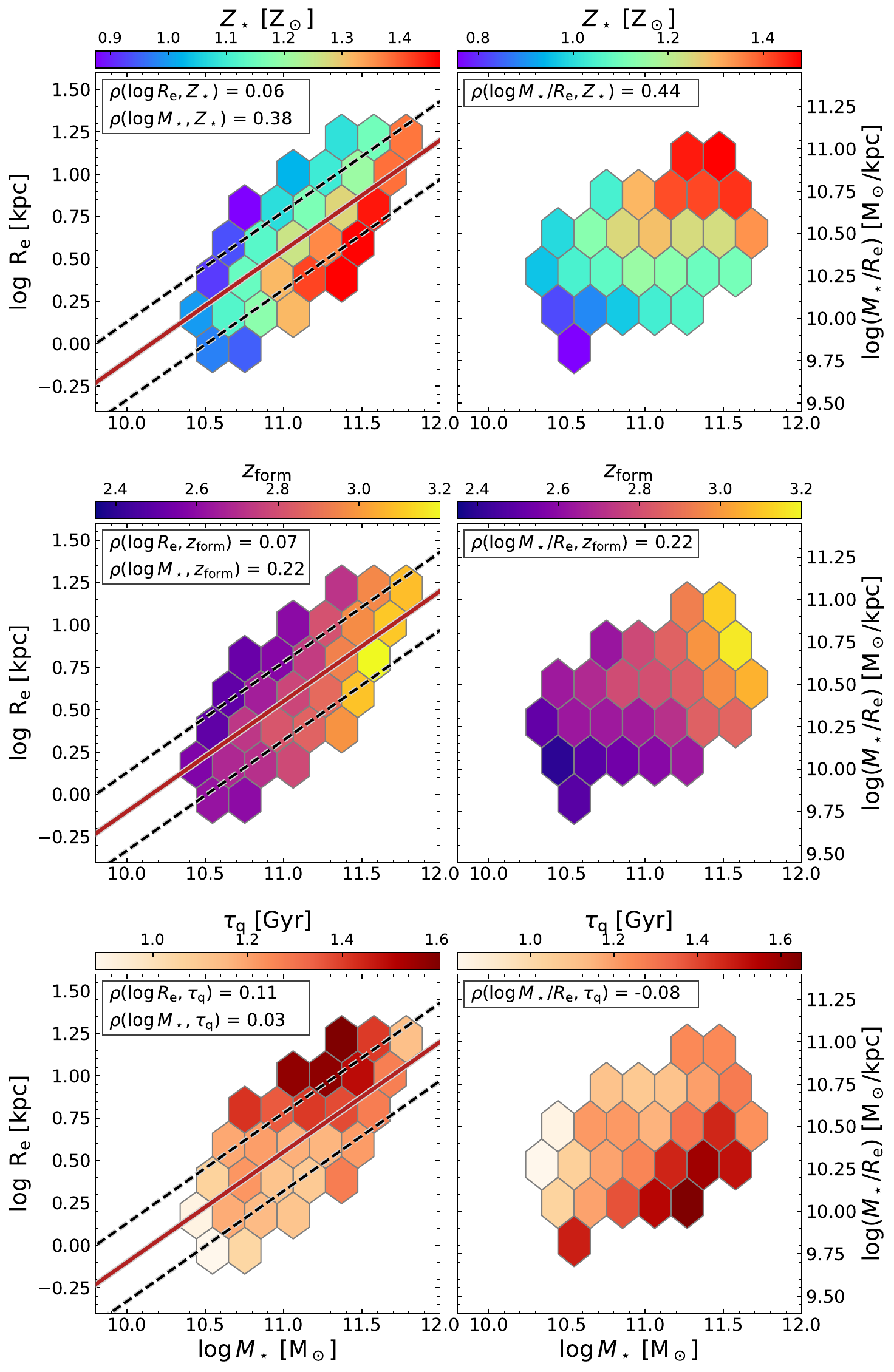}
    \caption{Mean stellar population properties and quenching timescales of quiescent galaxies on the size--mass plane (left), and $M_\star/R_\mathrm{e}$ versus $M_\star$ (right). Panels from top to bottom display: stellar metallicity ($Z_\star$), formation redshift ($z_\mathrm{form}$), and quenching timescale ($\tau_\mathrm{q}$). The solid red line is the best-fit size--mass relation for the quiescent population. The dashed black lines indicate the $1\sigma$ deviation from the best-fit relation. The stellar properties and quenching time, were averaged using the {\tt LOESS} method. Each hexbin contains at least five galaxies. In the top left corner of each panel we show the Spearman rank coefficients ($\rho$).}
    \label{fig:fig_7}
\end{figure}

Exploring the relationship between the size, stellar age, stellar metallicity, and quenching timescale could shed more light on the various evolutionary paths that galaxies follow, and possibly explain the observed scatter on the size--mass plane. The left column of Fig.~\ref{fig:fig_7} shows the average trends in stellar metallicity ($Z_\star$), formation redshift ($z_\mathrm{form}$),\footnote{The formation redshift, $z_\mathrm{form}$, is defined as the redshift corresponding to a lookback time equal to the mass-weighted age ($t_\mathrm{\star, mw})$). We computed $z_\mathrm{form}$ as: $z_\mathrm{form} = z(t_H(z_\mathrm{obs})-t_\mathrm{\star, mw})$, where $t_H(z_\mathrm{obs})$ is the age of the Universe at the galaxy’s spectroscopic redshift. This approach assumes that $t_\mathrm{\star, mw}$ closely traces the epoch of galaxy formation. Our estimated formation redshift values range from $z_\mathrm{form}=~$1--10.} and quenching timescale ($\tau_\mathrm{q}$) of the quiescent galaxy population on the size--mass plane. The quenching timescales were calculated from the individual SFHs of the quiescent and PSB galaxies. The average trends were derived with the Locally Weighted Regression ({\tt LOESS}) method \citep{Cleveland_doi:10.1080/01621459.1988.10478639} as implemented in the {\tt LOESS} routine\footnote{https://pypi.org/project/loess/} by \citet{Cappellari_2013MNRAS.432.1862C}. In particular, we used a linear local approximation, a smoothing factor of $f = 0.3$, and required that each hexbin contains at least five galaxies. The right column of Fig.~\ref{fig:fig_7} shows the correlation between the aforementioned properties with the mass-to-size ratio ($M_\star/R_\mathrm{e}$). We also estimated the Spearman rank coefficient ($\rho$) between the various quantities depicted in Fig.~\ref{fig:fig_7}. The values are shown in the top left corner of each panel.   

\begin{figure}[t]
    \centering
    \includegraphics[width=7cm]{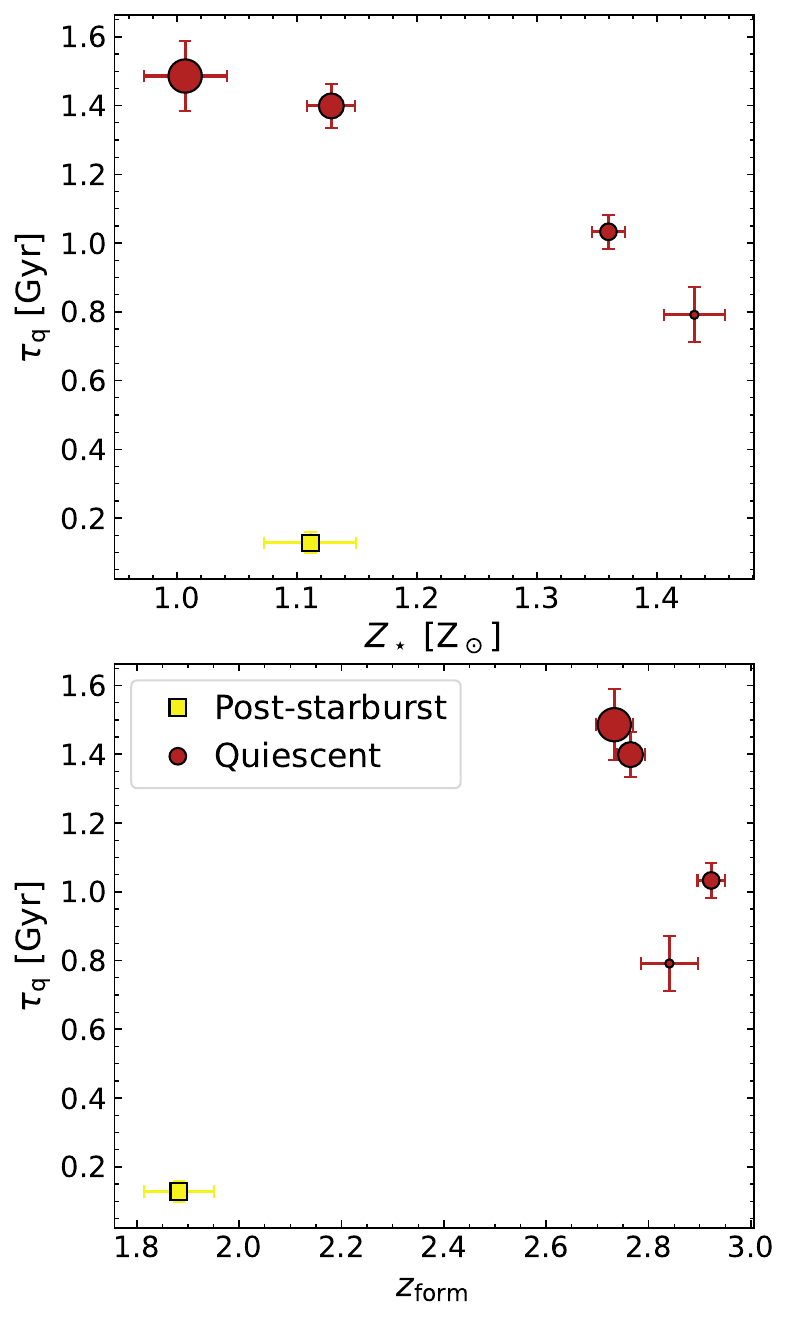}
    \caption{Median quenching timescales as a function of stellar metallicity (top), and formation redshift (bottom) for the quiescent galaxy population. We separated the data into four size bins: $\Delta \log (R_\mathrm{e}) \leq -0.23$, $-0.23 < \Delta \log (R_\mathrm{e}) < 0$, $0 < \Delta \log (R_\mathrm{e}) < 0.23$, and $\Delta \log (R_\mathrm{e}) \geq 0.23$. Larger circles correspond to larger galaxies. The yellow square marks the median value measured for the PSB population. The errorbars indicate the statistical uncertainty on the median ($\sigma/\sqrt{N}$). Larger quiescent galaxies quench more slowly, tend to be marginally younger, and have near-solar metallicities. The compact quiescent galaxies, are found to be metal-rich and seem to follow two distinct quenching pathways: one characterized by early and moderately fast quenching ($\langle \tau_\mathrm{q} \rangle = 0.8$~Gyr), and another characterized by rapid ($\langle \tau_\mathrm{q} \rangle = 0.13$~Gyr), late quenching.}
    \label{fig:fig_8}
\end{figure}

Examining the average trend in $Z_\star$, we find that galaxies below the size--mass relation tend to be metal-rich, consistent with their compact morphology and dense stellar populations \citep[e.g.,][]{Beverage_2021ApJ...917L...1B, Q1-SP044_Corcho_Caballero}. In contrast, galaxies above the relation exhibit a wider range of metallicities but, on average, have lower $Z_\star$, particularly at lower stellar masses. However, the Spearman rank coefficient between $R_\mathrm{e}$ and $Z_\star$ suggests a flat relation ($\rho = 0.06$). A much stronger correlation is observed between $Z_\star$ and $M_\star/R_\mathrm{e}$ ($\rho = 0.44$), indicating that quiescent galaxies with a stronger gravitational potential well ($\Phi \propto M_\star/R_\mathrm{e}$) retain a higher metallicity content \citep[e.g.,][]{Barone_2018ApJ...856...64B}. As $M_\star/R_\mathrm{e}$ is decreasing, we observe a decrease in $Z_\star$. This finding supports the idea that galaxy size plays a role in regulating the chemical composition of massive quiescent galaxies. Our results are consistent with \citet{Beverage_2021ApJ...917L...1B}, who analyzed 65 massive quiescent galaxies from the LEGA-C survey, demonstrating that chemical enrichment is governed by a galaxy's gravitational potential (as traced by $M_\star/R_\mathrm{e}$) \citep[see also][]{Zibetti_2022MNRAS.512.1415Z}. In \citet{Gallazzi_2025arXiv251207952G}, a similar correlation is shown between $Z_\star$ and stellar velocity dispersion ($\sigma_\star^2~\sim M_\star/R_\mathrm{e}$) at fixed $M_\star$, as well as a correlation with $M_\star$ at fixed $\sigma_\star$. Finally, \citet{Baker_2024MNRAS.534...30B} showed that $Z_\star$ in passive galaxies is primarily driven by $\sigma_\star$, concluding that black hole feedback plays a key role in quenching by suppressing gas inflows, enhancing $Z_\star$, and ultimately halting star formation.

The average trend with $z_\mathrm{form}$ suggests that quiescent galaxies $1\sigma$ (0.23~dex) above the size--mass relation formed 50\% of their stellar mass more recently than those $1\sigma$ below it, implying that more compact galaxies formed earlier. However, the overall correlation between $z_\mathrm{form}$ and size remains weak ($\rho = 0.07$). We also find a hint of a positive trend between $z_\mathrm{form}$ with $M_\star$ ($\rho = 0.22$). A stronger trend between $z_\mathrm{form}$ and $M_\star$ has been reported by \citet{Belli_2019ApJ...874...17B}, whereas \citet{Carnall_2019MNRAS.490..417C} and \citet{Tacchella_2022ApJ...926..134T} reported only a negligible dependence of $z_\mathrm{form}$ with stellar mass, in agreement with our results. Furthermore, $z_\mathrm{form}$ shows a weak positive correlation with $M_\star/R_\mathrm{e}$ ($\rho = 0.22$), indicating that galaxies with stronger gravitational potentials tend to form earlier (i.e., are older). For a sample of 625 quiescent galaxies from the SAMI Galaxy Survey \citep{Bryant_2015MNRAS.447.2857B}, \citet{Barone_2018ApJ...856...64B} found a moderate positive correlation between stellar population age and $M_\star/R_\mathrm{e}$ ($\rho = 0.39$). \citet{Beverage_2021ApJ...917L...1B} reported an opposite but similarly weak trend between stellar age and $M_\star/R_\mathrm{e}$ ($\rho = -0.1$), though their sample was limited to 65 LEGA-C galaxies. Based on \dfn~measurements, \citet{Wu_2018ApJ...868...37W} found that at a fixed stellar mass, smaller quiescent galaxies tend to be older, consistent with our results.

The general trends with $Z_\star$ and $z_\mathrm{form}$ shown in Fig.~\ref{fig:fig_7} remain consistent regardless of the method used to derive the stellar properties. We confirmed these trends using alternative estimates of the mass-weighted age and metallicity from both {\tt BaStA} \citep{Gallazzi_2025arXiv251207952G} and {\tt BAGPIPES} \citep{Kaushal_2024ApJ...961..118K}, consistently finding that relatively large quiescent galaxies tend to quench later and have lower stellar metallicities (see Appendix~\ref{apdx:B}).

At fixed stellar mass, we find that, on average, $\tau_\mathrm{q}$ increases with galaxy size, aligning with our results in Fig.~\ref{fig:fig_6}. Quiescent galaxies above or on the size--mass relation quench more slowly than those below the relation. However, we find a negligible correlation between size and quenching timescale ($\rho = 0.11$). Conversely, compact quiescent galaxies undergo a faster quenching than extended quiescent galaxies of the same mass (see also Appendix~\ref{apdx:C} for a comparison between the spectra of rapid, fast, and slow quenching galaxies of similar mass). We also find a weak negative correlation between $\tau_\mathrm{q}$ and $M_\star/R_\mathrm{e}$ ($\rho = -0.08$), where, at fixed mass, galaxies with stronger gravitational potentials quench more rapidly. 

In Figure~\ref{fig:fig_8}, we summarize the key trends in stellar metallicity, $z_\mathrm{form}$, quenching time, and galaxy size by plotting the median trends for the quiescent galaxies in four size bins based on the size metric $\Delta \log (R_\mathrm{e})$, as well as the PSB population. The median quenching timescales in each size bin were calculated based on the $\tau_\mathrm{q}$ values as measured from the individual SFHs of the quiescent population. From this figure, we can infer that  larger quiescent galaxies tend to have longer quenching timescales and near-solar metallicities, while the most compact systems quench faster and are metal-rich. In terms of $z_\mathrm{form}$, there is marginal evidence that compact quiescent galaxies formed earlier ($z_\mathrm{form}\simeq2.9$) than their more extended counterparts ($z_\mathrm{form}\simeq2.7$), corresponding to a time difference of about 200~Myr.

These trends supports the idea that galaxy size influences chemical enrichment \citep{Beverage_2021ApJ...917L...1B}. The observed change in metallicity may be driven by dry minor merger events, as the accretion of low-mass metal-poor satellites could decrease the metallicity of the progenitor galaxy \citep[e.g.,][]{Shankar_2010MNRAS.403..117S, Carollo_2013ApJ...773..112C}. In contrast, compact galaxies likely formed stars rapidly and early, leading to high metal enrichment via limited dilution from pristine gas, and strong AGN or supernova feedback, or intense starbursts, that can rapidly deplete the available cold gas within a few hundred Myr \citep{Zolotov_2015MNRAS.450.2327Z, Wu_2018ApJ...868...37W, Chauke_2019ApJ...877...48C, Wu_2020MNRAS.494.5636W}.

The median size of the 84 PSBs in our sample lies just below the size--mass relation, placing them in the second size bin (i.e., $-0.23 < \Delta \log (R_\mathrm{e}) < 0$). Based on their median properties, we find that PSBs undergo the most rapid quenching among the quiescent population, with a median quenching timescale of $\langle \tau_\mathrm{q} \rangle = 0.13\pm0.03$~Gyr. This short timescale aligns with both observational studies at similar redshifts \citep[e.g.,][]{Wild_2020MNRAS.494..529W} and cosmological hydrodynamic models, which suggest that rapid quenching can also be driven either by the mechanical expulsion of gas due to AGN activity \citep[e.g.,][]{Rodriguez_Montero_2019MNRAS.490.2139R} or by major mergers and multiple minor mergers \citep{Pawlik_2019NatAs...3..440P, Davis_2019MNRAS.484.2447D}. Additionally, \citet{Park_2023ApJ...953..119P} reported similarly a fast quenching timescale for 12 young quiescent galaxies at $z\sim1.5$, with an average transition time of $0.3$~Gyr. There is also ample evidence of fast quenching at higher redshifts, as highlighted in several recent studies \citep{Carnall_2023Natur.619..716C, Carnall_2024MNRAS.534..325C, de_Graaff_2024NatAs.tmp..284D, Urbano_Stawinski_2024MNRAS.528.5624U, Weibel_2025ApJ...983...11W, Turner_2025MNRAS.537.1826T}.

The median stellar metallicity of PSBs is higher than that of the largest quiescent galaxies, but lower than that of the most compact ones. The high metallicities of PSBs are consistent with recent findings from 45 local PSB galaxies in the MaNGA survey \citep{Leung_2024MNRAS.528.4029L}, which showed that most PSBs experienced significant metallicity enrichment during recent starbursts, supporting a merger-driven origin involving rapid quenching and efficient metal recycling. Finally, PSBs formed 50\% of their stellar mass significantly later (by $\sim1$~Gyr) than the rest of the quiescent galaxies in our sample. 

An interesting avenue for future work is to link our results on quenching timescales and galaxy sizes with chemical abundance patterns, particularly $\alpha$-enhancement. Upcoming capabilities in {\tt Prospector} will allow for self-consistent $\alpha$-enhanced models, which are expected to be crucial for deriving accurate physical properties of high-redshift galaxies \citep{Park_2025ApJ...994..165P}.

\section{Discussion} \label{sec:discussion}

\subsection{Multiple pathways to quiescence} \label{subsec:path_to_q}

Observational studies show that, at a fixed stellar mass, larger quiescent galaxies tend to be younger, as indicated by stellar age tracers such as \dfn~and \hd~\citep{Wu_2018ApJ...868...37W}. Several other studies have similarly reported that younger galaxies tend to be larger in size \citep[e.g.,][]{Ribeiro_2016A&A...593A..22R, Fagioli_2016ApJ...831..173F, Williams_2017ApJ...838...94W, Wu_2018ApJ...868...37W, Baes_2020A&A...641A.119B, Barone_2022MNRAS.512.3828B, Nersesian_2023A&A...673A..63N}, with size growth attributed to young stars forming at larger radii or minor mergers contributing to the outskirts \citep[e.g.,][]{van_Dokkum_2010ApJ...709.1018V, Vulcani_2014MNRAS.441.1340V, Shankar_2015ApJ...802...73S, Oogi_2016MNRAS.456..300O}. However, some studies suggest only a weak correlation between size growth and in situ star formation \citep[e.g.,][]{van_Dokkum_2013ApJ...771L..35V, van_der_Wel_2014ApJ...788...28V, van_Dokkum_2015ApJ...813...23V, Whitaker_2017ApJ...838...19W}. Regardless, galaxies evolve along different quenching pathways, undergoing morphological transformations in the process.

Cosmological simulations suggest that quenching timescale decreases with stellar mass. High-mass galaxies ($M_\star > 10^{11}~\mathrm{M}_\odot$) at $z\sim0$ quench within $\sim 1$~Gyr, while lower-mass galaxies ($M\star < 10^{10}~\mathrm{M}_\odot$) take 2--3 Gyr \citep{Nelson_2018MNRAS.475..624N, Park_2022MNRAS.515..213P}. Simulations also indicate that smaller star-forming galaxies quench earlier \citep{Genel_2018MNRAS.474.3976G}, while more extended massive galaxies at $z \sim 2$ quench later \citep{Gupta_2021ApJ...907...95G}. These discrepancies may be attributed to differences in the epoch of observations, with environmental quenching occurring at lower redshifts \citep[e.g.,][]{Moutard_2018MNRAS.479.2147M, Webb_2020MNRAS.498.5317W}. Observations show a very weak decreasing trend between quenching timescales and increasing stellar mass, with massive halos having shorter quenching timescales \citep[][albeit their sample of galaxies is limited and the uncertainties on the halo masses are large]{Tacchella_2022ApJ...926..134T}.

Our results reinforce the idea that galaxies follow a broad diversity of quenching pathways, ranging from rapid to slow quenching \citep[e.g.,][]{Barro_2016ApJ...820..120B, Wu_2018ApJ...868...37W, Moutard_2018MNRAS.479.2147M, Belli_2019ApJ...874...17B, Moutard_2020MNRAS.495.4237M, Tacchella_2022ApJ...926..134T}. The majority of quiescent galaxies quench on timescales of $\sim$1~Gyr or longer, significantly exceeding the typical dynamical time ($\sim$0.1~Gyr). This suggests that gradual processes such as gas depletion and strangulation dominate for most galaxies, producing minimal structural changes \citep{Leroy_2008AJ....136.2782L, Schiminovich_2010MNRAS.408..919S, Bigiel_2011ApJ...730L..13B, Saintonge_2011MNRAS.415...61S, Schruba_2011AJ....142...37S, Huang_2012ApJ...756..113H, Tacconi_2013ApJ...768...74T, Genzel_2015ApJ...800...20G, Tacconi_2018ApJ...853..179T, Tacconi_2020ARA&A..58..157T}. In contrast, only a subset of galaxies, such as PSBs, quench on very short timescales of $\lesssim$0.1--0.3~Gyr, consistent with the dynamical timescales of violent processes such as mergers, disk instabilities, or AGN-driven feedback, which can rapidly transform galaxies into compact remnants \citep{Zolotov_2015MNRAS.450.2327Z, Tacchella_2016MNRAS.458..242T}.

Based on the median trends in Fig.~\ref{fig:fig_8}, we find that compact quiescent galaxies are consistent with both early, moderately fast quenching and with more rapid, late quenching. While this may suggest the existence of multiple quenching channels, our data are also compatible with a continuous distribution of quenching timescales, as argued by \citet{Martin_2007ApJS..173..342M} in the local Universe. In either case, the results indicate that different physical mechanisms can drive quenching across galaxy populations, potentially producing similar morphological outcomes despite diverse evolutionary histories.

\subsection{Caveats} \label{subsec:caveats}

There are several sources of uncertainty that could potentially impact the conclusions of our analysis. One of these uncertainties is the definition of the transitioning and quiescence boundaries, that directly influences the calculation of $\tau_\mathrm{q}$. In this work, we adopted the boundaries from \citet{Tacchella_2022ApJ...926..134T}, who analyzed 161 massive quiescent galaxies at similar redshifts ($z_\mathrm{obs} \sim 0.8$). In Fig.~\ref{fig:fig_a1}, we show that these boundaries represent the LEGA-C sample exceptionally well. Moreover, \citet{Tacchella_2022ApJ...926..134T} found that this definition introduces no systematic bias in $\tau_\mathrm{q}$, except for a subset of rapidly quenching galaxies, where the timescales tend to be overestimated. While small shifts in the boundary definitions could lead to variations in the measured $\tau_\mathrm{q}$, we tested alternative thresholds (keeping the width of the transition zone fixed) and found that the resulting differences in $\tau_\mathrm{q}$ were typically around 0.15~Gyr. Therefore, the median quenching timescales remain largely unaffected, and we conclude that the adopted definition is appropriate for our analysis.

We caution that part of the apparent decline in the early star formation activity may be influenced by our SFH parameterization. In our non-parametric model, the SFR is assumed to be piecewise constant in lookback time bins, while the stellar mass grows monotonically. This naturally drives a decline in the inferred specific SFR with cosmic time, even if the underlying SFR is flat. As a result, the early decline seen in Fig.~\ref{fig:fig_4}--\ref{fig:fig_6} should not be overinterpreted as a robust physical trend. Rather, it reflects in part the modeling assumptions, and similar declines would appear under this parameterization regardless of the true early SFH. A detailed comparison of different SFH models would be required to disentangle physical trends from modeling artifacts.

This analysis also assumes that each SFH represents the evolution of a single galaxy. In reality, some of the quiescent galaxies in our sample may have experienced multiple merger events prior to observation. In such cases, the reconstructed SFH could reflect a composite history of multiple progenitors rather than a single evolutionary track. This blending can bias the measurement of key parameters like $\tau_\mathrm{q}$, potentially smoothing out signatures of rapid quenching. 

Another source of uncertainty arises from the measurement of mass-weighted stellar ages and metallicities. Within the LEGA-C team, these properties are derived by combining spectra and photometry using three SED fitting methods: {\tt BAGPIPES} \citep{Kaushal_2024ApJ...961..118K}, {\tt Prospector} \citep{Nersesian_2025A&A...695A..86N}, and {\tt BaStA} \citep{Gallazzi_2025arXiv251207952G}. A detailed comparison in \citet[][see their Appendix~E]{Gallazzi_2025arXiv251207952G} reveals notable discrepancies in stellar metallicity, with variations up to 0.2~dex for quiescent galaxies and 0.4~dex for star-forming ones. These differences partly stem from methodology: {\tt Prospector} and {\tt BAGPIPES} assume a constant metallicity over time, while {\tt BaStA} provides light-weighted averages. Differences in the range of metallicity values also contribute to the increased scatter among the three SED fitting methods. For quiescent galaxies, age estimates are more consistent (within $\sim 0.08$~dex) between {\tt BaStA} and {\tt Prospector}, though {\tt BAGPIPES} often clusters ages around $\log(\mathrm{Age}/\mathrm{yr}) \sim 9.5$. For star-forming galaxies, age estimates across methods show better agreement overall, though uncertainties from {\tt Prospector} and {\tt BAGPIPES} still appear underestimated. Despite the large one-to-one variations among the different fitting methods, the average trends we presented in this work are consistent (see Appendix~\ref{apdx:B}).

\section{Summary and conclusions} \label{sec:conclusions}

In this paper, we investigated the relationship between galaxy sizes, SFHs, quenching times, and stellar population properties by leveraging spectrophotometric modeling results from {\tt Prospector} \citep{Leja_2017ApJ...837..170L, Johnson_2021ApJS..254...22J} and the LEGA-C DR3 dataset \citep{Nersesian_2025A&A...695A..86N}. The main goal of our analysis was to determine whether there is a link between galaxy sizes, SFHs, and quenching times. From the trends of the median SFHs in Fig.~\ref{fig:fig_5}, and Fig.~\ref{fig:fig_6}, we report that:

\begin{itemize}

\item galaxies at redshift $0.6 < z < 1$ show a great diversity of SFHs, with galaxies either actively star-forming, transitioning, or quenching. We measure that the transition from star formation to quiescence lasted on average $\langle\tau_\mathrm{q}\rangle = 1.23\pm0.04$~Gyr. Spectroscopically selected, compact PSB galaxies quenched on average much more rapidly with $\langle\tau_\mathrm{q}\rangle = 0.13\pm0.03$~Gyr.

\item at a fixed size, the most massive star-forming galaxies transition earlier than the less massive ones. This observation aligns with the downsizing scenario, where more massive galaxies tend to quench earlier. At a fixed stellar mass, we find a subtle indication that larger galaxies sustain their star-formation activity for a longer duration compared to their smaller counterparts.

\item at a fixed stellar mass, quiescent galaxies lying $1\sigma$ (0.23~dex) below the size--mass relation quench faster than galaxies on or $1\sigma$ above the size--mass relation. At a fixed size, we do not find clear evidence of a strong trend between quenching timescale and stellar mass, and if such a trend exists, it is rather weak.

\end{itemize}

Based on the $\tau_\mathrm{q}$ measurements from the individual SFHs of quiescent galaxies, and the median trends of $\tau_\mathrm{q}$ as a function of galaxy size, stellar metallicity, and formation redshift (Fig.~\ref{fig:fig_7} and Fig.~\ref{fig:fig_8}), we find that:

\begin{itemize}

\item larger quiescent galaxies quench more slowly and have near-solar metallicities, while the most compact quiescent galaxies, having a strong gravitational potential, are found to be metal-rich and to quench significantly faster. This supports the idea that galaxy size can regulate the chemical enrichment of galaxies \citep{Beverage_2021ApJ...917L...1B}. The observed change in metallicity may be driven by dry minor merger events, as the accretion of low-mass metal-poor satellites could decrease the metallicity of the progenitor galaxy. In this case, longer $\tau_\mathrm{q}$ could simply reflect the mixing of different stellar population that quenched at different times, thus stretching the effective $\tau_\mathrm{q}$ we measured on the integrated population.

\item larger quiescent galaxies also tend to be marginally younger, while compact quiescent galaxies seem consistent with both early, moderately fast quenching and with more rapid, late quenching. While this may suggest the existence of multiple quenching channels, our data are also compatible with a continuous distribution of quenching timescales. These findings suggest that different physical mechanisms drive quenching across galaxy populations, potentially leading to similar morphological outcomes despite differing evolutionary histories.

\end{itemize}

\begin{acknowledgements} 
AN gratefully acknowledges the support of the Belgian Federal Science Policy Office (BELSPO) for the provision of financial support in the framework of the PRODEX Programme of the European Space Agency (ESA) under contract number 4000143347. PFW acknowledges funding through the National Science and Technology Council grants 113-2112-M-002-027-MY2. ARG acknowledges support from the INAF-Minigrant-2022 "LEGA-C" 1.05.12.04.01. This research made use of Astropy,\footnote{\url{http://www.astropy.org}} a community-developed core Python package for Astronomy \citep{Astropy_2013A&A...558A..33A, Astropy_2018AJ....156..123A}. 
\end{acknowledgements}



\bibliographystyle{aa}
\bibliography{References} 

\begin{appendix}
%

\onecolumn

\section{Quenching boundaries} \label{apdx:A}

\begin{figure*}[h!]
\includegraphics[width=\textwidth]{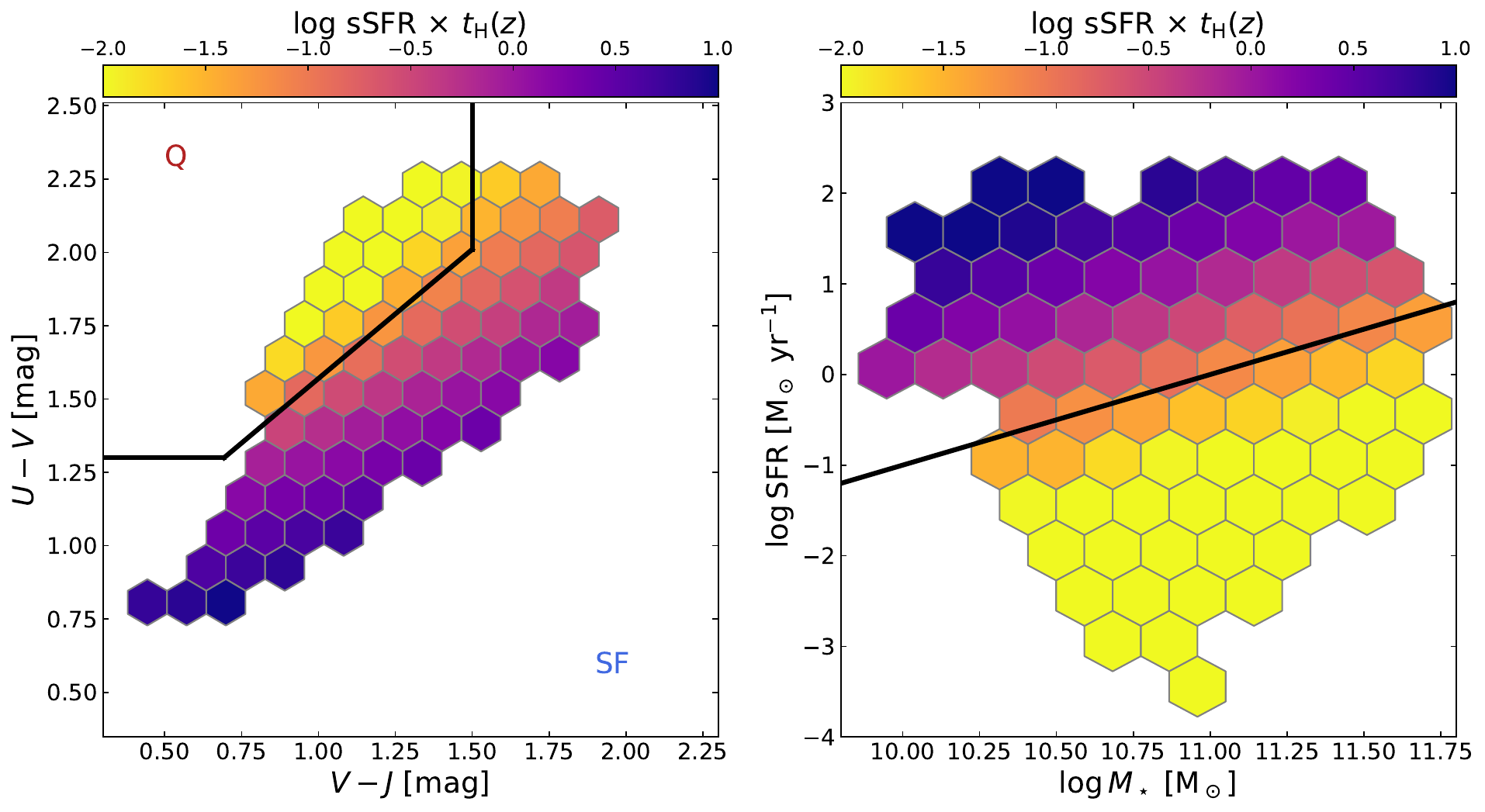}
     \caption{Distribution of the average $D$ of our sample of 2908 galaxies in the $UVJ$ (left) and SFR--$M_\star$ (right) planes. The average trends were derived with the Locally Weighted Regression ({\tt LOESS}) method \citep{Cleveland_doi:10.1080/01621459.1988.10478639, Cappellari_2013MNRAS.432.1862C}. The black lines indicate the separation between star formation and quiescence in the $UVJ$ diagram based on the definition of \citep{Muzzin_2013ApJS..206....8M}, and in the SFR--$M_\star$ plane ($\mathrm{sSFR} = 10^{-11}~\mathrm{yr}^{-1}$).}
     \label{fig:fig_a1}
\end{figure*}


One of the main results of our study is based on the definition of the mass-doubling number $D$, defined as:

\begin{equation}
    D(z) = \mathrm{sSFR}(z) \times t_\mathrm{H}(z),
\end{equation}

\noindent and representing the number of times the stellar mass doubles within the age of the Universe at redshift $z$, at the current sSFR \citep{Tacchella_2022ApJ...926..134T}. Based on the sample and properties of their sample, massive quenching galaxies at a median redshift of $z=0.8$, \citet{Tacchella_2022ApJ...926..134T} defined the quenching boundary to be between $1/20 < D(z) < 1/3$ or $-1.3 < \log D(z) < -0.48$. Figure~\ref{fig:fig_a1}, presents the $UVJ$ and SFR--$M_\star$ planes color-coded with the doubling number $D$ in $log$. Based on the average trends of $D$ in the $UVJ$ and SFR--$M_\star$ planes, the definition by \citet{Tacchella_2022ApJ...926..134T} seems to represent the LEGA-C galaxy sample well, and the cuts being meaningful. While small shifts in the boundary definitions could lead to variations in the measured $\tau_\mathrm{q}$, we tested alternative thresholds (keeping the width of the transition zone fixed) and found that the resulting differences in $\tau_\mathrm{q}$ were typically around 0.15~Gyr. Therefore, the median quenching timescales remain largely unaffected, and we conclude that the adopted definition is appropriate for our analysis.

\section{Validation with alternative models} \label{apdx:B}

\begin{figure*}[t]
    \centering
    \includegraphics[width=0.7\textwidth]{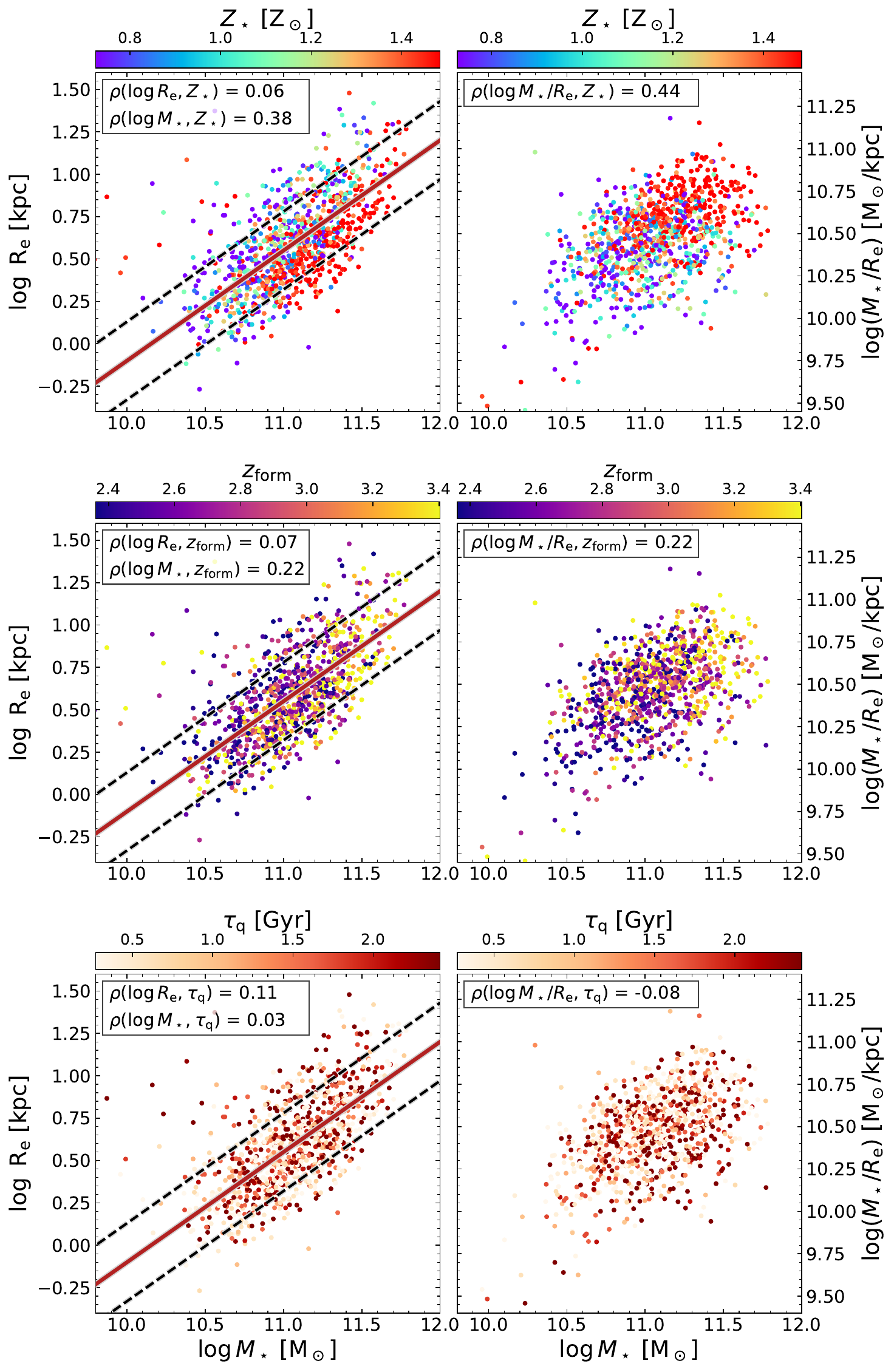}
    \caption{Scatter plots of the stellar population properties and quenching timescales of quiescent galaxies on the size--mass plane (left), and $M_\star/R_\mathrm{e}$ versus $M_\star$ (right). Panels from top to bottom display: stellar metallicity ($Z_\star$), formation redshift ($z_\mathrm{form}$), and quenching timescale ($\tau_\mathrm{q}$). The solid red line is the best-fit size--mass relation for the quiescent population. The dashed black lines indicate the $1\sigma$ deviation from the best-fit relation. The stellar properties and quenching time, were averaged using the {\tt LOESS} method. Each hexbin contains at least five galaxies. In the top left corner of each panel we show the Spearman rank coefficients ($\rho$).}
    \label{fig:fig_b1}
\end{figure*}

Here we present the actual scatter plots of the stellar population properties and quenching timescales of quiescent galaxies on the size--mass plane, and $M_\star/R_\mathrm{e}$ versus $M_\star$ (see Fig.~\ref{fig:fig_b1}). From these scatter plots, we can reach the same conclusions as from the average trends presented in Fig.~\ref{fig:fig_7}, especially for $Z_\star$ and $z_\mathrm{form}$. The scatter plot with $\tau_\mathrm{q}$ is less prominent, and the {\tt LOESS} method is necessary to recover the mean trends of the population.

\begin{figure*}
  \centering
  \subfigure[]{\includegraphics[width=0.48\textwidth]{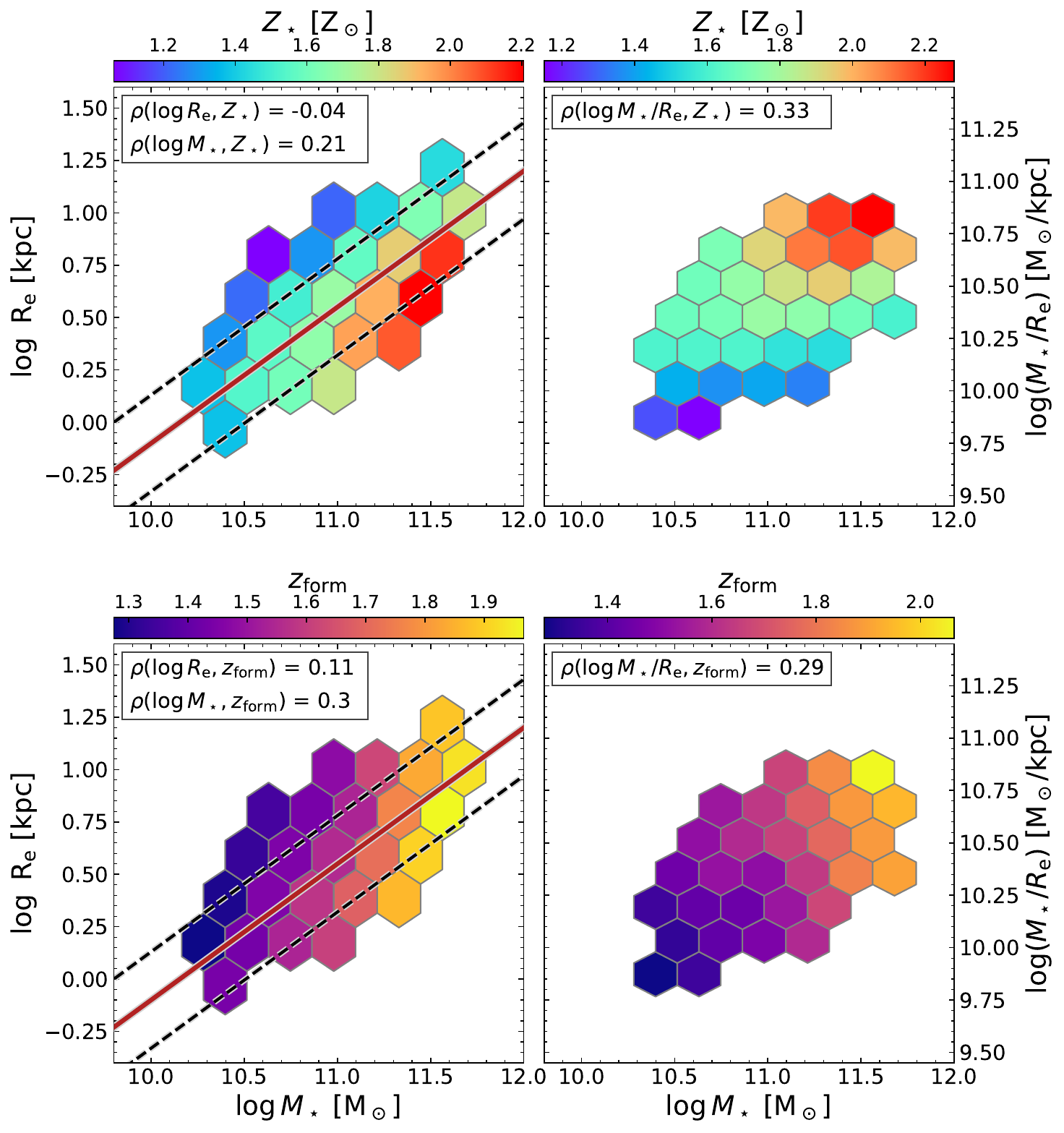}}
  \hfill
  \subfigure[]{\includegraphics[width=0.48\textwidth]{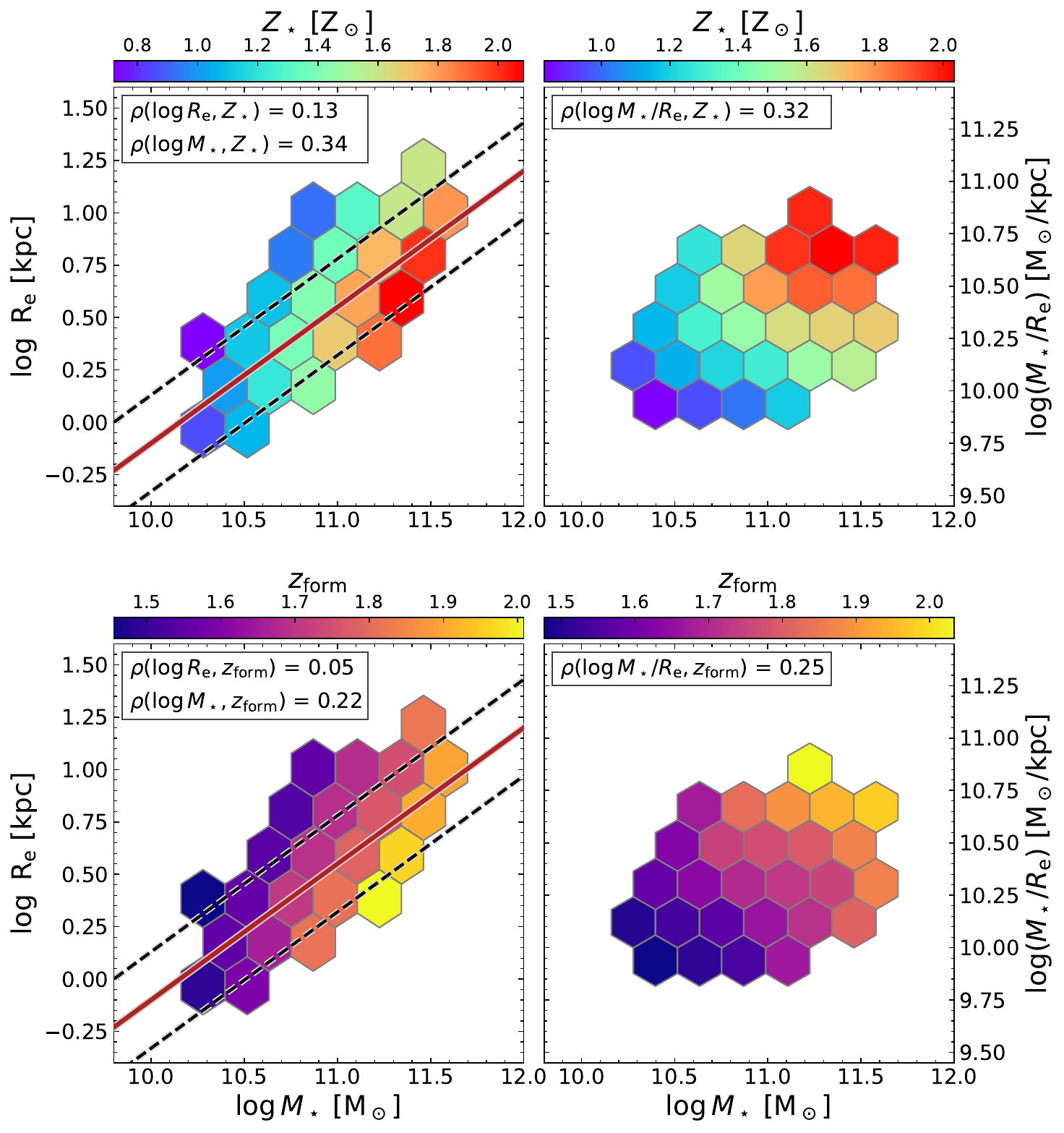}}
  \caption{Mean stellar population properties of quiescent galaxies on the size--mass plane, and $M_\star/R_\mathrm{e}$ versus $M_\star$. Panels from top to bottom display: stellar metallicity ($Z_\star$), and formation redshift ($z_\mathrm{form}$). The solid red line is the best-fit size--mass relation for the quiescent population. The dashed black lines indicate the $1\sigma$ deviation from the best-fit relation. The stellar properties and quenching time, were averaged using the {\tt LOESS} method. Each hexbin contains at least five galaxies. In the top left corner of each panel we show the Spearman rank coefficient ($\rho$). (a) Stellar properties from \citet{Gallazzi_2025arXiv251207952G} derived with {\tt BaStA} \citep{Gallazzi_2005MNRAS.362...41G, Zibetti_2017MNRAS.468.1902Z} (b) Stellar properties from \citet{Kaushal_2024ApJ...961..118K} derived with {\tt Bagpipes} \citep{Carnall_2018MNRAS.480.4379C} }
  \label{fig:fig_b2}
\end{figure*}

Moreover, within the LEGA-C team, we have applied three alternative methodologies to fit the LEGA-C DR3 spectra: {\tt Prospector} \citep[][this work]{Nersesian_2025A&A...695A..86N}, {\tt BaStA} \citep{Gallazzi_2025arXiv251207952G} and {\tt Bagpipes} \citep{Kaushal_2024ApJ...961..118K}. The {\tt BaStA} code \citep{Gallazzi_2005MNRAS.362...41G, Zibetti_2017MNRAS.468.1902Z, Zibetti_2022MNRAS.512.1415Z} incorporates stochastic parametric SFHs and derives the stellar properties using Lick index and broadband photometry modeling. In \citet{Kaushal_2024ApJ...961..118K}, we used the Bayesian SPS code {\tt Bagpipes} for a spectrophotometric analysis, similar in scope to this work. The key differences are that {\tt Prospector} employs a nonparametric SFH and marginalizes over emission lines, while {\tt Bagpipes} uses parametric SFHs without line marginalization. 

In Fig.~\ref{fig:fig_b2}, we present the same trends shown in Fig.~\ref{fig:fig_7}, but using the results from the two alternative SED fitting approaches. Despite differences in SED fitting methods, the overall trends with $Z_\star$ and $z_\mathrm{form}$ remain consistent across all methods, confirming the robustness of our findings. The Spearman correlation coefficients also agree closely with those reported in Fig.~\ref{fig:fig_7}. While some variation in absolute values is present, it is expected. Both {\tt BaStA} and {\tt Bagpipes} explore a broader metallicity range, allowing for higher $Z_\star$ values. Additionally, both rely on parametric SFHs, which tend to yield younger mass-weighted ages, and thus lower $z_\mathrm{form}$ compared to the nonparametric SFHs used in {\tt Prospector}, which is known to return older stellar ages \citep{Leja_2019ApJ...877..140L, Leja_2019ApJ...876....3L}.

\section{Comparing spectra of rapid, fast, and slow quenching galaxies} \label{apdx:C}

In this section, we want to validate the quenching timescales ($\tau_\mathrm{q}$) inferred from the reconstructed SFHs using {\tt Prospector}, based on the observed spectrophotometric data from the LEGA-C survey. To this end, we compare the observed and modeled spectra for a subset of galaxies selected based on their \hda~absorption strength and size metric, $\Delta \log (R_\mathrm{e})$. Specifically, galaxies are divided into PSB and quiescent populations according to their \hda~strength (see Sect.~\ref{subsec:psb}). The quiescent galaxies are further split into 'compact' [$\Delta \log (R_\mathrm{e}) \leq -0.23$] and 'extended' [$\Delta \log (R_\mathrm{e}) \geq 0.23$]. Based on the {\tt Prospector}-inferred quenching times, the median $\tau_\mathrm{q}$ values for the PSBs, compact quiescent, and extended quiescent galaxies are 0.13, 0.8, and 1.5~Gyr, respectively (see also Fig.~\ref{fig:fig_8}), corresponding to rapid, fast, and slow quenching regimes. The median stellar mass across all three bins is comparable, at $\log M_\star/\mathrm{M}_\odot = 11 \pm 0.2$.

\begin{figure}[t]
    \centering
    \includegraphics[width=\columnwidth]{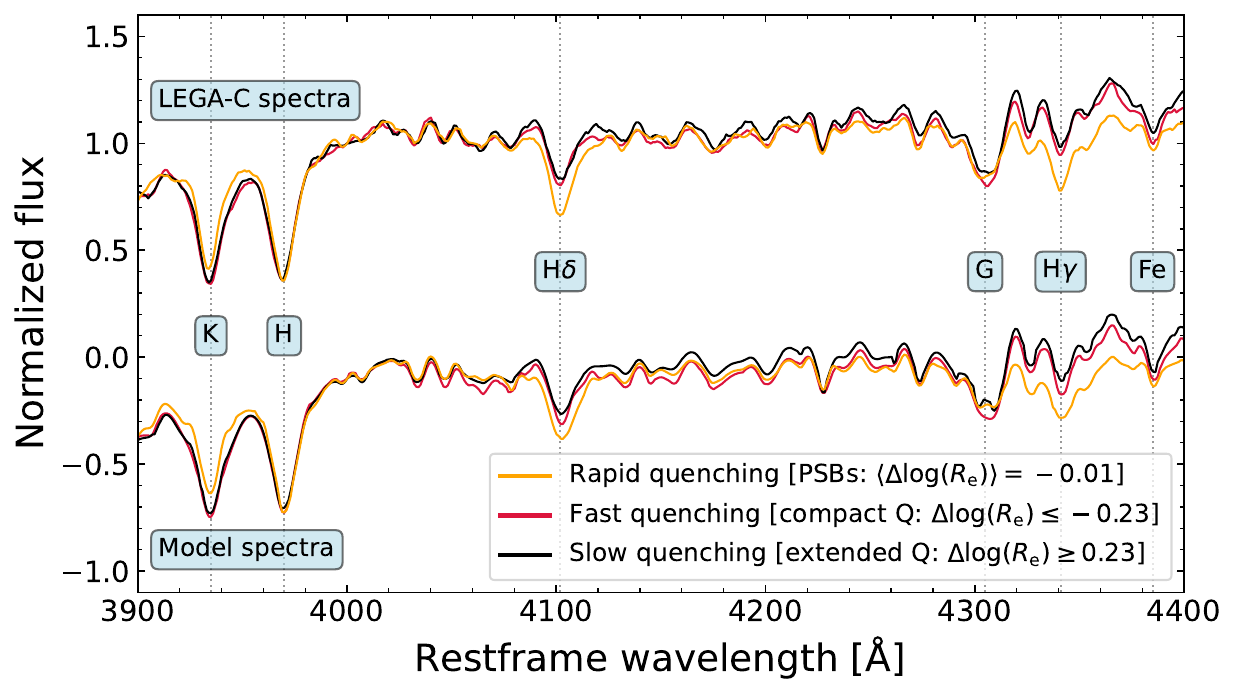}
    \caption{Median observed spectra of rapid (yellow), fast (red), and slow (black) quenching galaxies. Galaxies are split by \hda~strength, to PSBs and quiescent. Furthermore, the quiescent galaxies are split based on their size metric, $\Delta \log (R_\mathrm{e})$, to, small [$\Delta \log (R_\mathrm{e}) \leq -0.23$] and large [$\Delta \log (R_\mathrm{e}) \geq 0.23$]. The spectra are smoothed, and normalized at $4000~\AA$ for comparison, while the model spectra are shown with an arbitrary vertical offset for clarity. Several absorption features are marked with dotted lines.}
    \label{fig:fig_c1}
\end{figure}

Figure~\ref{fig:fig_c1} presents the median, smoothed, and normalized spectra for the three galaxy groups. Clear differences are visible in key absorption features, including the Balmer lines (\hd~and \hg), the Ca\ion{II}~K+H lines, and the G-band (CH absorption around $4300~\AA$). The PSBs show the strongest Balmer absorption, characteristic of a young, recently quenched stellar population. Compact quiescent galaxies also display relatively strong Balmer lines, consistent with a more rapid quenching history compared to their extended counterparts. They also exhibit a more pronounced G-band feature, indicating an older and metal-rich underlying stellar population, consistent with the marginally higher $z_\mathrm{form}$ and $Z_\star$ presented in Fig.~\ref{fig:fig_8}.

Compact quiescent galaxies also show stronger Fe absorption lines, suggesting higher stellar metallicities, again in agreement with the trends shown in Fig.~\ref{fig:fig_8}. In contrast, extended quiescent galaxies exhibit the weakest Balmer absorption, consistent with slow quenching. Their K feature strength is comparable to that of the compact group, suggesting similar intermediate-age stellar populations and supporting their similar formation redshifts. Interestingly, the G-band strength of extended quiescent galaxies is comparable to that of PSBs, potentially indicating residual star formation and a younger mass-weighted stellar age. Again, this is consistent with the slightly lower $z_\mathrm{form}$ inferred in Fig.~\ref{fig:fig_8}.

The spectral differences observed across the three galaxy groups, are consistent with the quenching timescales, metallicities, and formation redshifts presented in Fig.~\ref{fig:fig_8}. This consistency supports the reliability of the {\tt Prospector}-inferred SFHs and quenching times, validating the results of our analysis.   

\end{appendix}
\end{document}